\documentclass[aos, preprint]{imsart}

\RequirePackage[colorlinks,citecolor=blue,urlcolor=blue]{hyperref}
\usepackage[T1]{fontenc}
\usepackage{amsmath, bm}
\usepackage{amsthm}
\usepackage{float}
\usepackage{lscape}
\usepackage{amssymb}
\usepackage{tabularx}
\usepackage[dvips]{epsfig}
\usepackage[dvips]{graphics}
\usepackage{graphicx}
\usepackage{dsfont}
\usepackage{url}
\usepackage{slashbox}
\usepackage{natbib}
\usepackage{multirow}
\usepackage{verbatim}
\usepackage[dvipsnames]{xcolor}
\usepackage[normalem]{ulem}

\usepackage{lineno}

\allowdisplaybreaks

\newtheorem{theorem}{Theorem}[section]
\newtheorem{corollary}[theorem]{Corollary}
\newtheorem{lemma}[theorem]{Lemma}

\theoremstyle{definition}

\makeatletter
\newcommand*{\rom}[1]{\expandafter\@slowromancap\romannumeral #1@}
\makeatother

\newcommand{\phiHat}{\widehat{\varphi}_j}
\newcommand{\PhiHat}{\widehat{\Phi}_{j_1, \ldots, j_r}}
\newcommand{\PhiPop}{\Phi_{j_1, \ldots, j_r}}

\newcommand{\pnorm}[1]{\lVert #1 \rVert_{\mathfrak{p}}}

\newcommand{\PP}{\mathbb{P}}
\newcommand{\EE}{\mathbb{E}}
\newcommand{\RR}{\mathbb{R}}
\newcommand{\YY}{\mathbb{Y}}
\newcommand*{\vertbar}{\rule[-0.5ex]{0.5pt}{3.05ex}}

\definecolor{newMaterial}{rgb}{0.05,0.04, 0.075}

\startlocaldefs
\endlocaldefs

\begin{document}

\begin{frontmatter}

\title{Marginal and Interactive Feature Screening of Ultra-high Dimensional Feature Spaces with Multivariate Response}
\runtitle{Marginal and Interact. Screening Multivar. Response}


 \author{\fnms{Randall} \snm{Reese}\corref{}\ead[label=e1]{Randall.Reese@inl.gov}}\thanksref{t1}\thanksref{t2}
 \thankstext{t1}{Corresponding Author}
 \thankstext{t2}{Support for this work was provided in part by the Presidential Doctoral Research Fellowship at Utah State University.}
 \address{Idaho National Laboratory\\ P.O. Box 1625\\ Idaho Falls, ID 83415\\ \printead{e1}}
 \affiliation{Idaho National Laboratory}


\runauthor{Reese}
\begin{abstract}
 When the number of features exponentially outnumbers the number of samples, feature screening plays a pivotal role in reducing the dimension of the feature space and developing models based on such data. While most extant feature screening approaches are only applicable to data having univariate response, we propose a new method (GenCorr) that admits a multivariate response. Such an approach allows us to more appropriately model multiple responses as a single unit, rather than as unrelated entities, which avails more robust analyses in relation to complex traits embedded in the covariance structure of multiple responses. The GenCorr framework allows for the screening of both marginal as well as interactive features. It is demonstrated that GenCorr possesses the desirable property of strong sure screening. In the marginal case, we examine the superior numerical performance of GenCorr in comparison to two current methods for multivariate marginal screening via an assortment of empirical simulations. We also present several simulations inspecting GenCorr's performance in multivariate interaction screening. A culminating real data analysis demonstrates the performance of our method on GWAS data. 
\end{abstract}



\end{frontmatter}



\section{Introduction}\label{intro}
Variable selection and feature screening methods for high and ultrahigh dimensional data sets have been an oft explored topic in fields such as regression 
modelling, machine learning, and classification. Early approaches such as lasso \citep{Tibshirani}, SCAD \citep{SCAD2001}, adaptive-lasso \citep{ZouAdaptLasso2006}, and 
elastic net \citep{elasticNet} focused on penalized regularization regression. These methods were followed by various implementations of sure independence 
screening, as pioneered by \cite{FanLv:2008}. \cite[See also e.g. ][]{FanSamworthWu:2009, MMLE2010, ZhuLiLiZhu2011, DCSIScitation, Huang2014, 
CuiLiZhong2015}. 
While the sole or principal focus of these feature screening methods is the case when the response is univariate, we are interested here in developing a 
new approach for feature screening when the feature space is ultrahigh dimensional and the response is multivariate. The ability to feature screen 
ultrahigh dimensional feature spaces when the response is multivariate can allow us to develop more accurate classification and regression models because 
we can account for the covariance structure between the components of the response jointly.  The ideal multivariate screening method would possess the 
ability to screen for both marginal and interactive effects. 

Throughout this paper we will let $n$ represent the number of observations, $p$ represent the number of covariates, and $q$ represent the number of 
components in the response vector $\YY$. We will tacitly assume throughout this work that the feature space is ultrahigh dimensional in the classical 
sense of \cite{FanLv:2008}. 
Consider the multivariate regression model below:

\begin{equation*}
h(\YY) = {X}{B},
\end{equation*}
where $\YY \in \mathbb{R} ^ {n \times q}$ is the observed $n \times q$ matrix of responses, ${X} \in \mathbb{R} ^ {n \times p}$ {is the 
covariate or predictor matrix}, $B \in \mathbb{R} ^ {p \times q}$ is the parameter coefficient matrix, and $h$ is a link function. The 
$j^{th}$ row vector of $B$, $B_j$, is a vector of length $q$ corresponding to the coefficient vector of the $j^{th}$ 
predictor, where $j = 1,...,p$. For screening purposes, we will assume that the true model is sparse, with few predictors (relative to $p$) having a 
causative effect on the response. Our goal with this work is to introduce a feature screening method applicable to situations where $\YY$ is multivariate. 
Moreover, we will seek to not only establish our method as a viable option when screening for marginal effects in a multivariate response model, but also 
as the only existing method for \emph{interaction} effects screening when $q > 1$.

\cite{Sun2014kernel} present an approach to multivariate response modelling in the setting of classification using microarray data. While a tractable 
method for ultrahigh dimesional feature screening with respect to classification, the method is not generally applicable to broader multivariate response 
problems. Moreover, their techniques cannot be used to screen for interactions in any regard.  \cite{Ma2015testing} consider a feature screening process 
when the relationship between $q$ and $n$ is ultrahigh dimensional (but $p < n$). This situation is distinct (and opposite) from our assumptions here and 
will not be considered. Two well known methods for feature screening ultrahigh dimensional feature spaces for marginal effects when the response is 
multivariate are the SIRS method of \cite{ZhuLiLiZhu2011}, and the distance correlation (DC-SIS) method of \cite{DCSIScitation}. \citep[See also e.g., ]
[where SIRS and DC-SIS are put forth as the leading extant methods in marginal multivariate feature screening]{Liu2015, Chu2016feature}. It is important 
to note here, however, that neither SIRS nor DC-SIS allow for the screening of interactive effects. We will present below a new method for feature 
screening ultrahigh dimensional feature spaces with multivariate response.  Our method will be able to screen for both marginal effects and interactive 
effects. As a matter of comparison, we will compare the empirical results of our method in the marginal case with the results of SIRS and DC-SIS. 

The remainder of this paper will be organized as follows: In Section \ref{methods} we will present our new feature screening approach, along with the 
necessary notation and theoretical properties associated with this new method. This will be followed by Section \ref{simulations}, wherein we empirically 
compare our newly proposed method  with the existing approaches of SIRS and DC-SIS, as well as demonstrate the application of our method in the setting of 
interaction screening. Therein we will also present a real data analysis on a genome-wide-association-study (GWAS) data set for mice. A concluding 
discussion will be given in Section \ref{discussion}. Section \ref{proofs} will be devoted to proving the theoretical results of Section \ref{methods}.

\section{Feature Screening and Generalized Correlation}\label{methods}
Here we present an overview of a screening procedure built upon the concepts of the generalized correlation matrix. We first explore a novel process of 
screening for marginal effects of individual covariates on a multivariate response. This is then followed by the presentation of a yet hereunto unseen 
method for feature interaction screening when the response is multivariate. 

\subsection{Marginal Feature Screening Via Generalized Correlation}

Let $Z_1$ and $Z_2$ be univariate random variables. The linear (Pearson) correlation between $Z_1$ and $Z_2$ is given by
\[\varrho = \frac{\text{Cov}(Z_1, Z_2)}{\sqrt{\text{Var}(Z_1)\text{Var}(Z_2)}},\]
where the standard definitions of the variance and covariance are used. This concept of correlation can be generalized to greater than two variables in 
the following manner. 
Let $\bm{Z} = [Z_1, Z_2, Z_3, \ldots, Z_r]^T$ be an $r$-dimensional random variable with any distribution. 
Let $\Sigma$ be the standard variance-covariance matrix associated with $\bm{Z}$. We then can obtain a correlation matrix for $\bm{Z}$ as follows:
\[\text{corr}(\bm{Z}) = \left[\text{diag}\left({\Sigma}\right)\right]^{-1/2} {\Sigma}\left[\text{diag}\left({\Sigma}\right)\right]^{-1/2}.\]   
See Chapter 7 of \cite{KutnerLinReg2004} for a further discussion on the correlation matrix. 

This work with the correlation matrix and generalized correlation can be specifically extended to marginal effects in the following manner: For any one of the covariates, $X_j$, consider the correlation matrix of the random variable vector \[\mathcal{V}_j = [X_j, Y^{(1)}, Y^{(2)}, \ldots, Y^{(q)}],\]
where $Y^{(m)}$ is the $m^{th}$ component of $\YY$.  
The variance-covariance matrix of $\mathcal{V}_j$ can be written in block matrix form as below:
\renewcommand{\arraystretch}{2}%
\[
	\Sigma_{j} = \left(\begin{array}{@{\hspace{1.5em}}c|c@{\hspace{1.5em}}}
	\text{Var}(X_j) & \mathcal{C}^T\\ 
	\hline
	\mathcal{C} & \text{Cov}(\YY)
	\end{array}\right),
\]
\renewcommand{\arraystretch}{1.25}
where $\mathcal{C}$ is a $q$-vector whose $m$th entry is the covariance between $X_j$ and $Y^{(m)}$,
and $\text{Cov}(\YY)$ represents the variance-covariance matrix for the components of $\YY$. Let $\mathcal{H}_j$ be the generalized correlation for $\mathcal{V}_j$:
\[\mathcal{H}_j = \left[\text{diag}\left({\Sigma}_j\right)\right]^{-1/2} {\Sigma}_j\left[\text{diag}\left({\Sigma}_j\right)\right]^{-1/2}.\] 
This allows us to now construct a population quantity of a utility measure for covariate ranking. 
Let $\varphi_j = \lVert\mathcal{H}_j\rVert_{\mathfrak{p}},$  with $\lVert\cdot\rVert_{\mathfrak{p}}$ being any $\ell_{\mathfrak{p}}$ norm with $1 \leq \mathfrak{p} < \infty$. (Of note, $\lVert\cdot\rVert_{\infty}$ cannot be directly employed here as it will always be equal to one in this case). The $\mathfrak{p}$ used here is not to be confused with the $p$ used to denote the dimension of the covariate space. 
Herein we will examine empirical results under the taxi-cab (i.e. $\ell_1$) and Frobenius (i.e. $\ell_2$) matrix norms. Where necessary, these norms will be denoted by $\lVert\cdot\rVert_T$ and  $\lVert\cdot\rVert_F$ respectively. It should be noted, however, that any $\ell_{\mathfrak{p}}$-norm (with $\mathfrak{p} < \infty$) can be employed here from a theoretical standpoint. This will be reflected in the proofs (see Section \ref{proofs}) of the underlying theory to follow below.
The use of other matrix norms beyond the previously discussed entrywise $\ell_{\mathfrak{p}}$ norms (e.g. induced matrix norms, Schatten norms, $\ell_{\mathfrak{p}, \mathfrak{q}}$ norms, etc.) is reserved for a later work and extends beyond the scope of what we will examine in this paper.  
For further reference on the aforementioned taxi-cab and Frobenius norms, as well as any other $\ell_{\mathfrak{p}}$ and matrix norms, see e.g. \cite{Horn2012matrix}. 

We can create an estimator $\phiHat$ of each $\varphi_j$ as follows:
\begin{itemize}
\item The variance of each $X_j$ is estimated by \[\widehat{\text{Var}}(X_j) = \frac{1}{n-1} \sum_{i = 1}^n (X_{ij} - \overline{X}_j)^2,\]
where $\overline{X}_j$ is the sample mean of $X_j$. 
\item Each covariance in $\mathcal{V}$ is estimated by \[\widehat{\text{Cov}}(Y^{(m)}, X_{j}) = \frac{1}{n-1} \sum_{i = 1}^n \left(X_{ij} - \overline{X}_{j}\right)\left(Y^{(m)}_i - \overline{Y}^{(m)}\right),\]
where $\overline{X}_j$ is still the sample mean of $X_j$ and $\overline{Y}^{(m)}$ is the sample mean of $Y^{(m)}$. 
\item The covariance matrix of $\YY$ is to be estimated in the usual way, where
\[\widehat{\text{Cov}}\left(Y^{(\ell)}, Y^{(m)}\right) = \frac{1}{n-1} \sum_{i = 1}^n \left(Y^{(\ell)}_i - \overline{Y}^{(\ell)}\right)\left(Y^{(m)}_i - \overline{Y}^{(m)}\right).\]
\item This process ultimately results in the ability to create an estimator $\widehat{\Sigma}_{j}$ of the matrix ${\Sigma}_{j}$. We can then in turn create the following matrix:
\[\widehat{\mathcal{H}}_j = \left[\text{diag}\left(\widehat{\Sigma}_{j}\right)\right]^{-1/2} 
\widehat{\Sigma}_{j}\left[\text{diag}\left(\widehat{\Sigma}_{j}\right)\right]^{-1/2}.\] The matrix $\widehat{\mathcal{H}}_j$ is a natural estimator of the 
matrix $\mathcal{H}_j$ presented above. 
\item Application of the desired matrix norm to $\widehat{\mathcal{H}}_j$ in order to produce the resulting $\phiHat$ is a matter of routine calculation. 
Explicitly, we define
\[\phiHat = \pnorm{\widehat{\mathcal{H}}_j}.\]
\end{itemize}

Once $\phiHat$ is calculated for each $j = 1,2,\ldots, p$, we then rank all candidate predictors according to their associated $\phiHat$ value, from 
largest to smallest. Covariates associated with larger $\phiHat$ values are taken as having a larger association with the response $\YY$. Herein, we will 
call this newly proposed method GenCorr,
 in reference to the integral use of generalized correlation in the method. When reference to a specific norm is necessary, we 
will indicate as such. GenCorr-T will refer to instances where the taxi-cab norm is used. GenCorr-F will refer to instances where the Frobenius norm is 
used. 

\subsection{Extensions to Interaction Feature Screening}
From a purely theoretical vantage, this proposed method can be further extended to screen for $r$-way (with $r \geq 2$) interactions between predictors as 
follows:
\renewcommand{\arraystretch}{2}%
\[\text{Let }
	\Sigma_{j_1, j_2, \ldots, j_r} = \left(\begin{array}{@{\hspace{1.5em}}c|c@{\hspace{1.5em}}}
	\prod_{s = 1}^r \left[\text{Var}(X_j)\right] & \mathcal{K}^T\\ 
	\hline
	\mathcal{K} & \text{Cov}(\YY)
	\end{array}\right),
\]
\renewcommand{\arraystretch}{1.25}
where $\mathcal{K}$ is a $q$-vector whose $m$th entry is given by
\[\mathcal{K}_m = \kappa_{r+1}\left(Y^{(m)},~ X_{j_1},~ X_{j_2},~ \ldots,~ X_{j_r}\right), \quad m = 1,2,3, \ldots, q. \]
Here $k_{r+1}$ represents the $(r+1)$-way joint cumulant. For a further discussion on joint cumulants, see \cite{ReeseInteract}, \cite{NicaCum2006}, and \cite{SLJHuCum1991}. 
Note that by viewing the covariance of a component of $\YY$ and a given $X_j$ as the two-way joint cumulant, the previously outlined application of 
GenCorr to screen for marginal effects is really just a subcase of the method presented here for feature interaction.
As such, much of the relevant theory and proofs relating to GenCorr will be presented in the form of $r$-way interaction screening. The easily conceivable 
case where $r=1$ will account for the desired theoretical properties of GenCorr when applied to marginal effect screening. This ensuing presentation of GenCorr solely
via the broader framework of $r$-way interactions is done to avoid duplicating nearly identical theorems for marginal screening and interaction screening 
separately.  

As a matter of notation, define the following set of $r$-tuples with integer entries  
\[\mathcal{I} = \left\{(j_1, j_2, \ldots, j_r)~|~ 1 \leq j_1 < j_2 <\cdots < j_r \leq p \right\}.\]  
The set $\mathcal{I}$ contains all combinations of the covariate indices for which we can have an $r$-way interaction. 

Define the following:
\[\mathcal{H}_{j_1, j_2, \ldots, j_r} = \left(\left[\text{diag}\left(\Sigma_{j_1, j_2, \ldots, j_r}\right)\right]^{-1/2} \Sigma_{j_1, j_2, \ldots, j_r}\left[\text{diag}\left(\Sigma_{j_1, j_2, \ldots, j_r}\right)\right]^{-1/2}\right).\]
We can now create the following interaction utility measure for $r$-way covariate interaction ranking:
\[\text{Let } ~\Phi_{j_1, j_2, \ldots, j_r} =\lVert \mathcal{H}_{j_1, j_2, \ldots, j_r}  \rVert_{\mathfrak{p}},\]
where once again $\lVert \cdot  \rVert_{\mathfrak{p}}$ is any (entry-wise) $\ell_{\mathfrak{p}}$ norm with $\mathfrak{p} < \infty$.

As was the case with the construction of $\widehat{\varphi}_{j}$, it is a simple exercise to construct an estimator of $\Phi_{j_1, j_2, \ldots, j_r}$:
\begin{itemize}
\item The variance of each $X_j$ is estimated as before for $\widehat{\varphi}_{j}$. 
\item The $({r+1})$-way joint cumulant is estimated by 
\small\[\widehat{\kappa}_{r+1}(Y^{(m)}, X_{j_1}, \ldots, X_{j_r}) = \frac{1}{n} \sum_{i = 1}^n \left[~\prod_{s=1}^r\left(X_{ij_s} - \overline{X}_{j_s}\right)\right]\left(Y^{(m)}_i - \overline{Y}^{(m)}\right),\]
\normalsize
where $\overline{X}_{j_s}$ represents the sample mean of $X_{j_s}$ and $\overline{Y}^{(m)}$ is the sample mean of $Y^{(m)}$. 
\item The covariance matrix of $\YY$ is to be estimated in the usual way, as was done in the case of $\phiHat$. 
\item Quite like was done in the marginal effects screening case, one can now create an estimator $\widehat{\mathcal{H}}_{j_1, j_2, \ldots, j_r}$ of the matrix $\mathcal{H}_{j_1, j_2, \ldots, j_r}$. From this, the values of an estimator, denote it by $\PhiHat$, are a straight forward calculation. Explicitly, we define
\[\PhiHat = \pnorm{\widehat{\mathcal{H}}_{j_1, j_2, \ldots, j_r}}.\] 
\end{itemize}
After calculating $\PhiHat$ for each $r$-tuple in $\mathcal{I}$, we then can rank each interaction as most to least important based on the associated 
values of $\PhiHat.$ Larger $\PhiHat$ values are taken as having larger association with $\YY$. In this way, $\PhiHat$ will act as a screening utility for 
feature screening. 

In line with the concept of sparsity in ultrahigh dimensional feature screening, we can assume that only a small number of the interactions between 
covariates have a truly causative association with the response. 
Let $\mathcal{S}_F = \mathcal{I}$ represent what we will call the ``full model.'' This model is a model which admits every possible ($r$-way) interaction 
between the covariates in the feature space. Let $\mathcal{S} \subseteq \mathcal{S}_F$ denote an arbitrary model to be taken under examination. We will also 
define $X_{(\mathcal{S})}$ to be the set of all covariate interactions whose $r$-tuple indices are contained in $\mathcal{S}$.  Given a positive constant 
$c > 0$, we can define an estimated model:
\[\widehat{\mathcal{S}} = \{(j_1,\ldots, j_r) \in \mathcal{I} ~\mid~ \PhiHat > c\}.\] 
Here $\widehat{\mathcal{S}}$ represents a model selected by GenCorr given a predetermined cutoff $c>0$. Multiple approaches exist for determining this 
cutoff (see e.g. \cite{ZhuLiLiZhu2011}; \cite{ZHAO2012397}; \cite{Huang2014}; and \cite{KongWangWahba2015}).
 We suggest examining these methods for use with GenCorr, but will not explore the determination of a cutoff further. 

Let $\mathcal{D}\left({\YY} \mid {X_{(\mathcal{S})}}\right)$ represent the conditional distribution of $\YY$ given ${X_{(\mathcal{S})}}$. We will say that a model $\mathcal{S}$ is \emph{sufficient} if 
\[\mathcal{D}\left({\YY} \mid {X_{(\mathcal{S}_F)}}\right) = \mathcal{D}\left({\YY} \mid {X_{(\mathcal{S})}}\right).\]
The full model $\mathcal{S}_F$ is clearly sufficient. Ultimately, we are really only interested in finding the smallest (in terms of cardinality) 
sufficient model. The smallest sufficient model is also known as the \emph{true model}.  We will denote the true model by $\mathcal{S}_T$. We will denote 
the number of features in a given model by $|\mathcal{S}|$. (So the number of true or causative features would be written as $|\mathcal{S}_T|$). The 
principal aim in feature screening is producing an estimated model which not only \emph{contains} the true model, but moreover is the \emph{smallest} such 
model to contain all the true features (or, as the case may be, all true interactions).  

The most common situation where our approach to interaction screening can be applied in practice is in the case where $r = 2$
 (i.e. screening for two-way interactions).
  Computational limitations currently hedge what feasibly can be done with ultrahigh dimensional feature spaces when $r$ is greater than two. Owing to these limitations, all simulations pertaining to interaction screening presented herein (see Section \ref{simulations}, Simulation 5) will only deal with 
  two-way interactions.

\subsection{Theoretical properties}\label{thrtProp}
We first establish several conditions to facilitate the technical proofs that will be presented in Section \ref{proofs}.

\begin{enumerate}
    \item[(C1)] \emph{Bounds on the standard deviations}. We denote the variances of each $X_j$ and the variance of the components of $\YY$ as follows:  $\text{Var}(X_j) = \sigma_j^2$ and $\text{Var}(Y^{(m)}) = \sigma_{(m)}^2$. Assume that there exists a positive constant $\sigma_{\text{min}}$ such that for every $1\leq j \leq p$ and $1 \leq m \leq q$,
    \[
   		 \sigma_j,\sigma_{(m)} \geq \sigma_{min}.
    \]
    This allows us to exclude any covariates that are constant and thus have a standard deviation of {zero}. 
 
\vspace*{0.6cm} 
 
\item[(C2)] \emph{Entries of $~\mathcal{H}_{j_1, \ldots, j_r}$ for $(j_1, \ldots, j_r)$ in $\mathcal{S}_T$}. Assume that for each $(j_1, \ldots, j_r)$ in $\mathcal{S}_T$ there exists some positive constant $\omega_{j_1, \ldots, j_r} > 0$ and some $m$ in $\{1,2,\ldots, q\}$ such that 
\[\left|\frac{\kappa_{r+1}\left(Y^{(m)}, X_{j_1}, \ldots, X_{j_r}\right)}{\sigma_{(m)}\sigma_{j_1}\cdots\sigma_{j_r}}\right| > \omega_{j_1, \ldots, j_r}>0.\]  
This condition ensures that for every true or causative interaction, there is at least one component of $\YY$ with which the causative interaction has a non-zero association as determined by the joint cumulant. 

\vspace*{0.6cm}     
    
\item[(C3)] \emph{$\Phi_{j_1, \ldots, j_r}$ for $(j_1, \ldots, j_r)$ not in $\mathcal{S}_T$}. Define the following constant.
\[\gamma = (q+1) + \sum_{1\leq \ell, m \leq q} |\rho_{\ell m}|^2,\] where $\rho_{\ell m}$ is the correlation between $Y^{(\ell)}$ and $Y^{(m)}$. Note that $\gamma$ minimally equals $q+1$, meaning that $\gamma$ is necessarily positive. 
    We will assume that $\Phi_{j_1, \ldots, j_r} = \sqrt{\gamma}$ for any $(j_1, \ldots, j_r) \not\in \mathcal{S}_T$.     
\end{enumerate}

When some or all of these aforementioned conditions hold, we can state the following theorems.
\begin{theorem}{(Sure Screening)}\label{thm1}
Given that conditions (C1) and (C2) hold, there exists a positive constant $c > 0$ such that if 
\[\widehat{\mathcal{S}} = \{(j_1,\ldots, j_r) \in \mathcal{I} ~\mid~ \PhiHat > c\},\]
then we have 
\[\lim_{n \to \infty}\PP\left(\mathcal{S}_T \subseteq \widehat{\mathcal{S}}\right) = 1.\]
\end{theorem}
Note, however, that we have no guarantee in this case of $\mathcal{S}_T$ asymptotically containing the estimated model $\widehat{\mathcal{S}}$.

\begin{theorem}{(Strong Sure Screening)}\label{thm2}
Given that conditions (C1), (C2), and (C3) all hold, there exists a positive constant $c > 0$ such that if 
\[\widehat{\mathcal{S}} = \{(j_1,\ldots, j_r) \in \mathcal{I} ~\mid~ \PhiHat > c\},\]
then we have 
\[\lim_{n \to \infty}\PP\left(\mathcal{S}_T = \widehat{\mathcal{S}}\right) = 1.\]
\end{theorem}

This strong sure screening property given in Theorem \ref{thm2} is naturally more difficult to obtain than the (weak) sure screening property of Theorem \ref{thm1}. However, strong sure screening ensures that we not only obtain (asymptotically) an estimated model that \emph{contains} the true model, but furthermore that we obtain an estimated model that asymptotically \emph{equals} $\mathcal{S}_T$ with probability equal to one. 
Beyond the theorems on sure screening stated above, we can also obtain the following corollaries.
\begin{corollary}\label{cor1}
There exists a value $\Phi_{\min}$ such that \[\PhiPop > \Phi_{\min} > \sqrt{\gamma} > 0\] whenever $(j_1, \ldots, j_r) \in \mathcal{S}_T$. 
\end{corollary}

\begin{corollary}\label{cor2}
The estimator $\PhiHat$ converges \emph{uniformly} in probability to $\Phi_{j_1, \ldots, j_r}$. That is to say, for any $\varepsilon > 0$, 
\[\lim_{n \to \infty} \PP\left(\max_{(j_1,\ldots, j_r) \in \mathcal{I}}\left|\PhiHat - \Phi_{j_1, \ldots, j_r}\right| > \varepsilon \right) = 0.\]
\end{corollary}
These corollaries will come as a natural result of the proofs of Theorems \ref{thm1} and \ref{thm2} (See Section \ref{proofs}). 

\section{Simulations and Empirical Data Analysis}\label{simulations}
We will first evaluate the empirical performance of GenCorr via a collection of five groups of simulations. These simulations are then followed by a real 
data analysis on data from a genome-wide association mapping of outbred mice. The first four groups of simulations pertain to screening for marginal 
effects. We will compare the performance of GenCorr with two other methods for marginal effects screening when the response is multivariate: Sure 
Independence Ranking Screening (SIRS) \citep{ZhuLiLiZhu2011} and Distance Correaltion (DC-SIS) \citep{DCSIScitation}. Two distinct matrix norms for 
GenCorr will also be considered. We will denote GenCorr under the taxicab norm as GenCorr-T; GenCorr under the Frobenius norm will be denoted by GenCorr-
F.

\subsection{Simulation 1}
For this simulation we wish to examine the ability of GenCorr, SIRS, and DC-SIS to successfully screen for a small set of causative or ``true'' predictors in a linear model with normally distributed predictors. This simulation will be split into three parts. Each part will explore the effect of different methods for generating the coefficients for the causative predictors. 

\subsubsection{Simulation 1.A}
Here we take a sample size of $n = 60$ of each of $p = 3000$ predictors. 
First we generate each covariate, $X_j$ as follows:
\[X_j \sim \text{N}(0, \sigma = 5), \quad\text{sampled 60 times.}\]
Let $\Sigma$ be a $6 \times 6$ matrix given by \[\Sigma = \left[0.5^{|\ell-m|}\right]_{m, \ell}.\]
Next we create the following matrix $B$:
\[\text{For } k = 1,2,3 \text{ sample }(\beta_{k1}, \beta_{k2},\beta_{k3},\beta_{k4},\beta_{k5},\beta_{k6}) \sim \text{MVN}(\bm{0}, \Sigma), \]
\[\text{and let } B = \begin{pmatrix}
\beta_{11} & \beta_{12} & \beta_{13} & \beta_{14} & \beta_{15} & \beta_{16}\\
\beta_{21} & \beta_{22} & \beta_{23} & \beta_{24} & \beta_{25} & \beta_{26}\\
\beta_{31} & \beta_{32} & \beta_{33} & \beta_{34} & \beta_{35} & \beta_{36}
\end{pmatrix}.\]
We then construct the values of $\YY$ with $q = 6$ as follows:
\[\YY = \begin{pmatrix}\vertbar & \vertbar & \vertbar&\vertbar & \vertbar &  \vertbar \\
Y^{(1)}&Y^{(2)}& Y^{(3)}&Y^{(4)}&Y^{(5)}&Y^{(6)}\\
\vertbar & \vertbar & \vertbar&\vertbar & \vertbar &  \vertbar \end{pmatrix} 
= \begin{pmatrix}
\vertbar & \vertbar & \vertbar\\
X_1 & X_2 & X_3\\
\vertbar & \vertbar & \vertbar
\end{pmatrix}B.\]

Written more explicitly, \[Y^{(m)} = \beta_{1m}X_1 + \beta_{2m}X_2 + \beta_{3m}X_3.\]

This construction will mean that $X_1$, $X_2$ and $X_3$ are to be considered as causative, while the remaining $X_j$ will be taken as noise. We ran 400 replications of this simulation.

We first report the means of the best, medial, and worst rankings of the three causative predictors by each method in question. In doing this, we are not 
concerned with tracking the individual ranks of $X_1$, $X_2$, and $X_3$, but rather we focus our efforts on recording the ranks of the best, medial, and 
worst ranked true predictors, irrespective of which causative covariate was which. This allows us to gain overall insight into the minimal model size 
required (on average) to positively include, respectively, one, two, or all three of the causative predictors. The average worst ranking for a method 
informs us of the average minimum model size that would be necessary to assure all three of the true predictors are present.
Hence, a mean worst rank of 46 would mean that, on average, we would need a final model size of 46 to guarantee that we have included all three of the 
true predictors in our model. A perfect score of best, medial, and worst rankings would be (1,2,3), which corresponds to a minimum model size of three. By 
reporting each of the best, medial, and worst rankings, we can observe not only how large of a model we must on average have, but also how small of a 
model we can have and still retain one or more of the causative variables.  The mean rank results for each of the four methods (GenCorr-T, GenCorr-F, 
SIRS, DC-SIS) are given in Table \ref{table:Sim1}. 


Both SIRS and DC-SIS are observably inadequate as a screening method in this setting, as neither is able to obtain any semblance of returning a viable 
result consistently. However, under the taxicab norm, our GenCorr method performs admirably here, with an average worst ranking (i.e. average minimum 
necessary model size) just over 21. When the Frobenius norm is used, the results are even more impressive, with a minimum required model size just under 
ten. In either case, this means that GenCorr has successfully reduced the feature space to a collection with dimension less than $n = 60$.  

It can be beneficial to also examine the median of the best, medial, and worst rankings. (See Table \ref{table:Sim1Med}). From this we see that SIRS is 
actually able to determine some of  the true features with acceptably high accuracy at least half of the time. However, because SIRS seems especially 
vulnerable to error when the sample size is small (relative to $p$), the method struggles to produce consistent results on average. (Median worst ranks of 
3 for both applications of GenCorr compared to a median worst rank of 423.50 for SIRS).%

\subsubsection{Simulation 1.B}
We again employ a sample size of $n = 60$ for each of $p = 3000$ predictors. 
First we generate each covariate, $X_j$ as follows:
\[X_j \sim \text{N}(3, \sigma = 1), \quad\text{sampled 60 times.}\]
As before, let $\Sigma$ be a $6 \times 6$ matrix given by \[\Sigma = \left[0.5^{|\ell-m|}\right]_{m, \ell}.\]
The rows of the coefficient matrix $B$ are given as follows:
\[\text{For } k = 1,2,3 \text{ sample }(\beta_{k1}, \beta_{k2},\beta_{k3},\beta_{k4},\beta_{k5},\beta_{k6}) \sim \text{MVN}(\bm{3}, \Sigma). \]
Thus, as opposed to part A of the simulation, the coefficients are now generated from the multivariate normal distribution with means three instead of zero. Tables \ref{table:Sim1} and \ref{table:Sim1Med} report the mean and median rankings for Simulation 1.B. 

With the adjustment in the means of the predictors and their coefficients, we see an increased accuracy over part A on the part of SIRS (mean worst rank of 57.81 versus a mean worst rank of 890.155 in part A). Both implementations of GenCorr again yield the smallest minimum required model sizes on average (35.5425 and 22.65 for the taxicab and Frobenius norms, respectively). While the performance of DC-SIS improves under the current settings, it lags well behind GenCorr in terms of both mean and median minimum required model size.

\subsubsection{Simulation 1.C}
For this simulation, we set $n$ equal to 120 and let $p$ be 1500. This change in sample size is done to better avoid the vulnerabilities of SIRS under small (relative to $p$) sample size. Here we set $q$ equal to four. As was done in part A, we let $X_j \sim N(0, \sigma= 5)$. 

Based on Example 1 of DC-SIS (who in turn follow \cite{FanLv:2008}), for $m$ from 1 to 4 (inclusive) and $j = 1,2,3$ we define

\small
\[\beta_{jm} = (-1)^{U}(a + |Z|), \quad \text{where } a = \frac{4\log(n)}{\sqrt{n}}, ~ U\sim \text{Bernoulli}(0.4), ~\text{and } Z \sim N(0,1).\]

\normalsize
As before, we let 
\[Y^{(m)} = \beta_{1m}X_1 + \beta_{2m}X_2 + \beta_{3m}X_3.\]
Tables \ref{table:Sim1} and \ref{table:Sim1Med} depict the mean and median ranks of the three methods on question under these settings.

This method for constructing the coefficients of the causative features bears perfect scores for GenCorr (under both of the norms used). While both SIRS and DC-SIS improve upon their results from part A, these two methods lag far behind GenCorr in overall accuracy. (GenCorr achieves nearly perfect scores under both matrix norms; SIRS and DC-SIS obtain mean minimum model sizes over 100 times larger than perfect acquisition). 

\begin{table}[H]
\caption{Simulation 1 Mean Ranks}
\label{table:Sim1}
\vspace*{0.1cm}
\centering
\begin{tabular}{|c c c c| } 
\hline
\multicolumn{4}{|c|}{\textbf{Part A}}\\\hline\hline
& Best Rank & Medial Rank & Worst Rank\\ 
 \hline
  GenCorr-T& 1.0000 & 2.4000 &21.3025 \\
  GenCorr-F &1.0000 & 2.0300 & 9.715\\
  SIRS & 185.4550  &  397.4375&    890.1550\\ 
  DC-SIS &22.8325  &  296.1950&   1236.3275  \\\hline\hline
\multicolumn{4}{|c|}{\textbf{Part B}}\\\hline\hline
& Best Rank & Medial Rank & Worst Rank\\ 
 \hline
  GenCorr-T& 1.0000&    2.1650   &   35.5425  \\ 
  GenCorr-F& 1.0000& 2.0925       & 22.6500 \\
  SIRS &    1.0000  &    2.2950  &   57.8100\\ 
  DC-SIS &     2.4450   &  32.8375  &  342.5025  \\\hline\hline
\multicolumn{4}{|c|}{\textbf{Part C}}\\\hline\hline
& Best Rank & Medial Rank & Worst Rank\\ 
 \hline
 GenCorr-T& 1.0000  &  2.0000  &   3.0000 \\ 
 GenCorr-F& 1.0000 &2.0000& 3.0050\\
  SIRS &   66.2025  &  169.6350 &   327.8850 \\ 
  DC-SIS &10.6100   & 102.1875 &   522.7250  \\\hline
\end{tabular}
\end{table}

\begin{table}[H]
\caption{Simulation 1 Median Ranks}
\label{table:Sim1Med}
\vspace*{0.1cm}
\centering
\begin{tabular}{ |c c c c| } 
\hline
\multicolumn{4}{|c|}{\textbf{Part A}}\\\hline\hline
& Best Rank & Medial Rank & Worst Rank\\ 
 \hline
  GenCorr-T& 1.00 & 2.00 &3.00 \\ 
  GenCorr-F& 1.00 & 2.00 &3.00 \\
  SIRS & 1.00 &  19.50 &   423.50\\ 
  DC-SIS &1.00   &    39.00&      1098.00   \\\hline\hline
\multicolumn{4}{|c|}{\textbf{Part B}}\\\hline\hline
& Best Rank & Medial Rank & Worst Rank\\ 
 \hline
  GenCorr-T& 1.00 & 2.00 &4.00 \\
  GenCorr-F& 1.00 & 2.00 &3.00 \\ 
  SIRS & 1.00 &  2.00 &   5.00\\ 
  DC-SIS &1.00   &    3.00&      52.00   \\\hline\hline
\multicolumn{4}{|c|}{\textbf{Part C}}\\\hline\hline
& Best Rank & Medial Rank & Worst Rank\\ 
 \hline
  GenCorr-T& 1.00 & 2.00 &3.00 \\ 
  GenCorr-F& 1.00 &2.00 &3.00 \\
  SIRS & 1.00 &  2.00 &   9.00\\ 
  DC-SIS &1.00   &    3.00&      294.00   \\\hline
\end{tabular}
\end{table}

\subsection{Simulation 2}
Simulation 2 is constructed quite similarly to Simulation 1, however we now test the ability of each of GenCorr, SIRS, and DC-SIS to detect an \textit{exponential} relationship between $\YY$ and $X_1,$ $X_2$ , and $X_3$. We still retain $n = 60$, $p = 3000$, and $q = 6$ from before. 

\subsubsection{Simulation 2.A}
With each $X_j$ being constructed using the approach from Simulation 1.A, now define $\YY$ as follows:
\[Y^{(m)} = \text{exp}\bigg\{\beta_{1m}X_1 + \beta_{2m}X_2 + \beta_{3m}X_3\bigg\},\]
with the matrix $B$ being created in the same manner as found in Simulation 1.A. Once again we run 400 replications of this simulation.

The results for Simulation 2.A are given in Table \ref{table:Sim2A}.

\begin{table}[H]
\caption{Simulation 2 Mean Ranks}
\label{table:Sim2A}
\vspace*{0.1cm}
\centering
\begin{tabular}{ |c c c c| } 
\hline
& Best Rank & Medial Rank & Worst Rank\\ 
 \hline\hline
  GenCorr-T&  9.3100 &    97.2650  &  545.0075 \\
  	GenCorr-F	&8.7075 &    77.8800 &   493.4925\\
  SIRS & 194.2475 &   443.1475  &  921.8525 \\ 
  DC-SIS &464.4850  & 1134.8150&   2016.2150 \\\hline
\end{tabular}
\end{table}  
  
We also report in Table \ref{table:Sim2AMed} the median of each of the best, medial, and worst ranks.   

\begin{table}[H]
\caption{Simulation 2 Median Ranks}
\label{table:Sim2AMed}
\vspace*{0.1cm}
\centering
\begin{tabular}{ |c c c c| } 
\hline
& Best Rank & Medial Rank & Worst Rank\\ 
 \hline\hline
  GenCorr-T& 2.00  &     36.00 &      324.50   \\
  GenCorr-F& 3.00 & 31.00 &239.00 \\
  SIRS & 1.00     &   17.00&       383.00 \\ 
  DC-SIS &283.50  &    1025.00 &     2162.00\\\hline
\end{tabular}
\end{table}  
While none of the methods produce breathtaking results, GenCorr (under both the taxicab and Frobenius norms) is consistently the superior method in terms 
of producing the smallest required mean minimum model size (GenCorr achieves mean minimum model sizes that are about half that achieved by SIRS, and about 
four times smaller than the same for DC-SIS). Here we must keep in mind that, at times, feature screening is not so much concerned with the matter of 
selecting the true predictors, but rather removing those predictors which are definitively unimportant.
In cases such as this, the aim often is not to directly determine the smallest sufficient model, but rather to screen the set of $p$-many predictors into 
a collection of reduced size. (See, for example, the discussions in \cite{fan2010selective} and \cite{FanHanLiu:BigData} on this topic).  Penalized 
regression methods such as those mentioned in Section \ref{intro} can then be applied to the reduced set of predictors.

Because SIRS is only focused on ranking the covariates based their empirical cumulative distribution function, any monotonicly increasing transformation 
(e.g. an exponetional transformation) of the covariates will produce the same feature screening results as obtained from the original, untransformed, 
features (up to choice of random seed).   To that end, it should be noted that the results for SIRS in Simulation 1.A will differ from the results for 
SIRS here, as different random seeds were used to generate the simulation data sets in Simulation 1 and Simulation 2. This will be the case in Simulations 
3 and 4 as well. 

\subsection{Simulation 3}
The previous simulations explored the ability of GenCorr to successfully screen for causative covariates when both the response and the covariates are 
continuous. We now turn our attention to the screening of data with discrete predictors and discrete multivariate response.

\subsubsection{Simulation 3.A}
Here we use the same general setup as used in Simulation 1.A. The one change we implement is with the distribution of the predictors:
\[X_j \sim \text{Pois}(\lambda = 2).\] Outside of this change, we retain $n = 60$, $p = 3000$, $q = 6$, and perform 400 replications as before. The 
coefficient matrix $B$ is generated in the same fashion as was done in Simulation 1.A. The overall simulation model will thus be given by

\[\YY = \begin{pmatrix}\vertbar & \vertbar & \vertbar&\vertbar & \vertbar &  \vertbar \\
Y^{(1)}&Y^{(2)}& Y^{(3)}&Y^{(4)}&Y^{(5)}&Y^{(6)}\\
\vertbar & \vertbar & \vertbar&\vertbar & \vertbar &  \vertbar \end{pmatrix} 
= \begin{pmatrix}
\vertbar & \vertbar & \vertbar\\
X_1 & X_2 & X_3\\
\vertbar & \vertbar & \vertbar
\end{pmatrix}B.\]

This will mean that once again, $X_1$, $X_2$ and $X_3$ will be considered causative. 
The results for Simulation 3.A are given in Table \ref{table:Sim3} (mean ranks) and Table \ref{table:Sim3Med} (median ranks).

The results for GenCorr are highly encouraging. On average, the required minimum model size produced by GenCorr-T is just below 27, a reduction to a model 
size less than half of the sample size of 60. GenCorr-F produces even more favorable results, with a minimum required model size just above 15. Moreover, 
96.25\% of the replicates of GenCorr-F produce a required minimum  model size less than 60. By comparison, SIRS only obtains a required minimum model size 
less than 60 in 29.5\% of replicates. 
 
\subsubsection{Simulation 3.B}
Part B of Simulation 3 uses the same set up that part B of Simulation 1 used, with the one difference being the distribution of the covariates. Like 
Simulation 3.A, we let 
\[X_j \sim \text{Pois}(\lambda = 2).\]

The causative covariates will again be $X_1$, $X_2$, and $X_3$. 
The results for Simulation 3.B are given in Table \ref{table:Sim3} (mean ranks) and Table \ref{table:Sim3Med} (median ranks).

On average, GenCorr-T requires a minimum model size about ten to eleven features smaller than that required by SIRS to capture all causative features. 
GenCorr-F bests all methods, with a mean minimum model size just greater than 25. Under the settings of part B, DC-SIS improves drastically over its 
performance in part A. However, in spite of this improvement, DC-SIS still lags significantly behind GenCorr and SIRS.  

\subsubsection{Simulation 3.C}
The same overall setup as was used in Simulation 1.C is used here, yet we now take
 \[X_j \sim \text{Pois}(\lambda = 2).\]
 The values of $n = 120$, $p = 1500$, $q = 4$ and $400$ replicates are carried over from Simulation 1.C. 
We report the results for Simulation 3.C in Table \ref{table:Sim3} (mean ranks) and Table \ref{table:Sim3Med} (median ranks).

Both versions of GenCorr yield perfect $(1,2,3)$ scores in each of the 400 replication. SIRS and DC-SIS produce significantly less favorable results, 
especially when viewed in the light of mean minimum model size. Although the median ranks for these latter methods are more in line with the GenCorr 
results, the inconsistencies in the average case are concerning. 

\begin{table}[H]
\caption{Simulation 3 Mean Ranks}
\label{table:Sim3}
\vspace*{0.1cm}
\centering
\begin{tabular}{|c c c c| } 
\hline
\multicolumn{4}{|c|}{\textbf{Part A}}\\\hline\hline
& Best Rank & Medial Rank & Worst Rank\\ 
 \hline
  GenCorr-T&  1.0000  &    2.2375&     26.8025 \\ 
  GenCorr-F &1.0000  &    2.1850  &   15.3075\\
  SIRS & 186.3225  &  432.5900 &   837.1225\\ 
  DC-SIS &46.3400    &412.4975   &1440.4675 \\\hline\hline
\multicolumn{4}{|c|}{\textbf{Part B}}\\\hline\hline
& Best Rank & Medial Rank & Worst Rank\\ 
 \hline
  GenCorr-T&  1.0025 &     2.2000&     35.6125      \\
  GenCorr-F& 1.0025 & 2.0850&25.3350 \\
  SIRS &       1.0050  &    2.3150 &    46.4925 \\
  DC-SIS &1.0450   &   8.1075  &   146.0100\\\hline\hline
\multicolumn{4}{|c|}{\textbf{Part C}}\\\hline\hline
& Best Rank & Medial Rank & Worst Rank\\ 
 \hline
  GenCorr-T&  1.0000   &   2.0000  &    3.0000      \\
  GenCorr-F& 1.0000 & 2.0000 &3.0000 \\
  SIRS &     54.7700  &  140.9275 &   260.7725 \\
  DC-SIS & 5.5000 &    73.9400 &   466.0725  \\\hline
\end{tabular}
\end{table}

\begin{table}[H]
\caption{Simulation 3 Median Ranks}
\label{table:Sim3Med}
\vspace*{0.1cm}
\centering
\begin{tabular}{ |c c c c| } 
\hline
\multicolumn{4}{|c|}{\textbf{Part A}}\\\hline\hline
& Best Rank & Medial Rank & Worst Rank\\ 
 \hline
  GenCorr-T&  1.0   &      2.0 &        3.0  \\ 
  GenCorr-F& 1.0 & 2.0 & 3.0\\
  SIRS &    1.0  &      18.5  &      310.5\\ 
  DC-SIS & 3.0    &   106.5 &     1280.0   \\\hline\hline
\multicolumn{4}{|c|}{\textbf{Part B}}\\\hline\hline
& Best Rank & Medial Rank & Worst Rank\\ 
 \hline
  GenCorr-T&  1.0   &      2.0 &        4.0  \\ 
  GenCorr-F& 1.0 & 2.0 &3.0 \\
  SIRS &    1.0  &      2.0  &      5.0\\ 
  DC-SIS & 1.0    &   2.0 &     9.0 \\\hline\hline
\multicolumn{4}{|c|}{\textbf{Part C}}\\\hline\hline
  & Best Rank & Medial Rank & Worst Rank\\ 
 \hline
  GenCorr-T&  1.00   &      2.00 &        3.00  \\ 
  GenCorr-F& 1.00 & 2.00 &3.00 \\
  SIRS &    1.00  &      2.00  &      6.00\\ 
  DC-SIS & 1.00    &   6.00 &     176.50 \\\hline
\end{tabular}
\end{table}

\subsection{Simulation 4}
We now turn our attention to a model exhibiting an exponential relationship between the response and the causative covariates. 

\subsubsection{Simulation 4.A} 
As was the theme in each part of Simulation 3, we follow the overall setup of a previous simulation, yet with each covariate coming from the Poisson distribution with mean two. Here we emulate the approach used in Simulation 2.A, with the single difference being \[X_j \sim \text{Pois}(\lambda = 2).\] 
The values of $n = 60$, $p = 3000$, and $q = 6$ are retained from Simulation 2.A. We perform 400 replications. 
The average rank results for Simulation 4.A are given in Table \ref{table:Sim4A}. The median rank results are given in Table \ref{table:Sim4AMed}

\begin{table}[H]
\caption{Simulation 4 Mean Ranks}
\label{table:Sim4A}
\vspace*{0.1cm}
\centering
\begin{tabular}{ |c c c c| } 
\hline
& Best Rank & Medial Rank & Worst Rank\\ 
 \hline\hline
  GenCorr-T&  1.4350  &   16.3325 &   140.0025 \\ 
  GenCorr-F & 1.265   &   11.445 &    112.070\\
  SIRS &  195.7550  &  431.3475 &   832.6675\\ 
  DC-SIS &252.5000 &   908.6000 &  1899.4725 \\\hline
\end{tabular}
\end{table}  

\begin{table}[H]
\caption{Simulation 4 Median Ranks}
\label{table:Sim4AMed}
\vspace*{0.1cm}
\centering
\begin{tabular}{ |c c c c| } 
\hline
& Best Rank & Medial Rank & Worst Rank\\ 
 \hline\hline
  GenCorr-T&   1.00     &    2.00 &       16.00     \\ 
  GenCorr-F& 1.00 & 2.00 &14.00 \\
  SIRS &        1.00    &    18.50 &      317.50 \\ 
  DC-SIS &79.00   &    739.00 &     2022.50 \\\hline
\end{tabular}
\end{table}  

As was the case with Simulation 2.A, the mean minimum model sizes obtain by each method exceed the number of samples $n$. However, this is once again a 
case where the ultimate goal is not to singularly obtain a final model, but rather to reduce the feature set preparatory to performing further feature 
space reduction methods such as penalized regression. In both the mean and median rank cases, GenCorr looks far more promising in this regard than the 
other two methods. 
Of note, GenCorr-F finds at least two of the causative covariates to be within the top 30 most important covariates in 92.75\% of the replicates (GenCorr-
T follows closely at 90.25\%).  
SIRS obtains such results only 53.75\% of the time; DC-SIS obtains such in only 4\% of the replicates. 

\subsection{Simulation 5}
We now turn our attention to empirically examining the claim of GenCorr to not only screen for marginal effect of individual predictors on a multivariate 
response, but also the ability of the extended version of GenCorr to screen for \textit{interactive effects} on a multivariate response. We will explore 
results under both the taxi-cab norm and the Frobenius norm.

\subsubsection{Simulation 5.A}
Here we take a sample of $n = 100$ of each of $p = 1000$ predictors. 
First we generate each covariate as follows:
\[X_j \sim \text{N}(0, \sigma = 2), \quad\text{sampled 100 times.}\]
Let $B$ be the $2 \times 4$ matrix defined as follows:
\[B = \begin{pmatrix}
1 & -1 & -2.5 & 2\\ 2 & 1.5 & -2& 1
\end{pmatrix}.\]

We also define an $n \times q$ error matrix \[ E = \begin{pmatrix}
\vertbar & \vertbar & \vertbar&\vertbar  \\
\epsilon_{1},& \epsilon_2 & \epsilon_3 & \epsilon_4\\
\vertbar & \vertbar & \vertbar&\vertbar
\end{pmatrix}, \text{ with each $\epsilon_m \stackrel{i.i.d}{\sim} \text{MVN}(\bm{0}, I_n)$.}\]  

For $q = 4$, we then construct the values of $\YY$:
\[\YY = \begin{pmatrix}\vertbar & \vertbar & \vertbar&\vertbar  \\
Y^{(1)}&Y^{(2)}& Y^{(3)}&Y^{(4)}\\
\vertbar & \vertbar & \vertbar&\vertbar  \end{pmatrix} 
= \begin{pmatrix}
\vertbar & \vertbar\\
X_1X_2 & X_3X_4\\
\vertbar & \vertbar 
\end{pmatrix}B + E,\]
where $X_{j_1}X_{j_2}$ represents the product of $X_{j_1}$ and $X_{j_2}$. 


This construction will mean that the interaction between $X_1$ and $X_2$, as well as the interaction between $X_3$ and $X_4$, are to be considered as 
causative. All other pairwise interactions will act as noise.  For the time being, we will omit any causative marginal effects from the model. A model 
that also includes causative marginal effects will be considered in part C of this simulation. We ran 400 replications of this simulation.

The proportion of replicates for which each individual causative interaction is within the top five interactions is represented by $\mathcal{P}_{j_1,j_2}.
$ We will denote by $\mathcal{P}_{\text{top}}$ the proportion of replicates for which one of the true interactions is found to be \textit{the} most 
important interaction. The proportion of replicates for which \emph{both} causative interactions are determined to be within the top five most important 
interactions is given under $\mathcal{P}_{\text{a}}$. 
These proportions are given in Table \ref{table:Sim5_P}.  
We then also report the 25\%, 50\%, 75\% and 90\% quantiles of the required number of interactions to contain the true interactions.
These results are given in Table \ref{table:Sim5_Q}.

The values for both $\mathcal{P}_{1,2}$ and $\mathcal{P}_{3,4}$ are highly encouraging. Moreover, the $\mathcal{P}_{\text{top}}$ value tells us that in 
all but one of the replications GenCorr determines one of the true interactions to be the interaction most strongly associated with the response. (In the 
one replicate where neither true interaction was found to be the top interaction, the interaction between $X_1$ and $X_2$ was nevertheless found to be the 
\textit{second} most important interaction). Under the taxi-cab norm, GenCorr results in both causative interactions being found in the top five most 
important interaction 93.75\% of the time. GenCorr with the Frobenius norm improves upon this, finding both causative interactions to be within the top 
five most important interactions 95.50\% of the time. 
 The quantile values for part A in Table \ref{table:Sim5_Q} indicate that in a significant majority of cases, GenCorr (under either norm) can locate the 
 two true causative interactions with a high level of accuracy. 

\subsubsection{Simulation 5.B}
In this part of Simulation 5, we again use a sample size of $n=100$ and let $p = 1000$. Like with part A, we will be examining a model without causative 
marginal effects. However, unlike in part A, where the model coefficients were fixed for each replicate, we now will generate the coefficients on the true 
interactions anew for each of 400 replications of the simulation. Once again, each covariate is sampled from the normal distribution centered at zero and 
having standard deviation of two:
\[X_j \sim \text{N}(0, \sigma = 2).\]
Let $\Sigma$ be a $4 \times 4$ matrix defined by \[\Sigma = \left[0.5^{|\ell-m|}\right]_{m, \ell}.\]
Next we create the coefficient matrix $B$:
\[\text{For } k = 1,2 \text{ sample }(\beta_{k1}, \beta_{k2},\beta_{k3},\beta_{k4}) \sim \text{MVN}(\bm{3}, \Sigma), \]
\[\text{and let } B = \begin{pmatrix}
\beta_{11} & \beta_{12} & \beta_{13} & \beta_{14} \\
\beta_{21} & \beta_{22} & \beta_{23} & \beta_{24} \\
\end{pmatrix}.\]
The values of $\YY$ with $q = 4$ are the constructed as follows:
\[\YY = \begin{pmatrix}\vertbar & \vertbar & \vertbar &  \vertbar \\
Y^{(1)}&Y^{(2)}& Y^{(3)}&Y^{(4)}\\
\vertbar&\vertbar & \vertbar &  \vertbar \end{pmatrix} 
= \begin{pmatrix}
\vertbar & \vertbar \\
X_1 X_2 & X_3X_4\\
\vertbar & \vertbar 
\end{pmatrix}B.\]

Written more explicitly, this model is \[Y^{(m)} = \beta_{1m}X_1X_2 + \beta_{2m}X_3X_4.\]

This construction will mean that the interaction between $X_1$ and $X_2$ and the interaction between $X_3$ and $X_4$ will be considered as causative. All remaining pairwise interactions will be seen as noise. 

Like in part A, we report $\mathcal{P}_{j_1, j_2}$, $\mathcal{P}_{\text{top}}$, $\mathcal{P}_{\text{a}}$, $q_{25}$, $q_{50}$, $q_{75}$, and $q_{90}$. These results are given in Tables \ref{table:Sim5_P} (selection proportions) and \ref{table:Sim5_Q} (quantiles).

Even under the additional challenge of handling varying coeffcients on the causative interactions, GenCorr yet obtains consistently accurate results for 
both the taxi-cab and the Frobenius norms. These results strengthen our trust in GenCorr as a feature interaction screening method when the response is 
multivariate. It should be noted that although the $\mathcal{P}_{\text{a}}$ values hover just above 0.6, it is easy to confirm by use of the well known 
addition rule in probability that (outside of a single replication), whenever one of the causative interactions is not found to be within the top five 
most important interactions, the remaining causative interaction exercises a strong effect on the response and is positively identified in the top five 
most important predictors. Such an event occurs when one of the causative interactions dominates over the other causative interaction (likely due to 
random selection of a coefficient vector with comparatively small values being assigned to the non-dominant interaction).  

\subsubsection{Simulation 5.C}

For this part of Simulation 5, we will once again use a sample size of $n=100$ and let $p = 1000$. As always, we will run 400 replications of this part of 
the simulation. Once again, each covariate is sampled from the normal distribution centered at zero and having standard deviation of two:
\[X_j \sim \text{N}(0, \sigma = 2).\]

Unlike parts A and B, however, we now incorporate marginal effects into our model.  This is done as follows. 

Let $\Sigma$ be a $4 \times 4$ matrix defined by \[\Sigma = \left[0.5^{|\ell-m|}\right]_{m, \ell}.\]
Next we create the coefficient matrix $B$:
\[\text{For } k = 1,2,3,4,5,6 \text{ sample }(\beta_{k1}, \beta_{k2},\beta_{k3},\beta_{k4}) \sim \text{MVN}(\bm{3}, \Sigma), \]
\[\text{and let } B = \begin{pmatrix}
\beta_{11} & \beta_{12} & \beta_{13} & \beta_{14} \\
\beta_{21} & \beta_{22} & \beta_{23} & \beta_{24} \\
\beta_{31} & \beta_{32} & \beta_{33} & \beta_{34} \\
\beta_{41} & \beta_{42} & \beta_{43} & \beta_{44} \\
3\beta_{51} & 3\beta_{52} & 3\beta_{53} & 3\beta_{54} \\
3\beta_{61} & 3\beta_{62} & 3\beta_{63} & 3\beta_{64} \\
\end{pmatrix}.\]
The values of $\YY$ with $q = 4$ are then constructed as follows:
\[\YY = \begin{pmatrix}\vertbar & \vertbar & \vertbar &  \vertbar \\
Y^{(1)}&Y^{(2)}& Y^{(3)}&Y^{(4)}\\
\vertbar&\vertbar & \vertbar &  \vertbar \end{pmatrix} 
= \begin{pmatrix}
\vertbar &\vertbar&\vertbar &\vertbar&\vertbar & \vertbar \\
X_1& X_2 & X_3& X_4& X_1 X_2 & X_3X_4\\
\vertbar & \vertbar&\vertbar &\vertbar&\vertbar &\vertbar 
\end{pmatrix}B.\]

Component-wise for each component of $\mathbb{Y}$, we have \[Y^{(m)} = \beta_{1m}X_1 + \beta_{2m}X_2 +\beta_{3m}X_3 + \beta_{4m}X_4 + 3\beta_{5m}X_1X_2 + 3\beta_{6m}X_3X_4.\]

This construction will again allow the interaction between $X_1$ and $X_2$ and the interaction between $X_3$ and $X_4$ to be taken as causative.  
The addition of the marginal effects of $X_1$, $X_2$, $X_3$, and $X_4$ means that GenCorr will now be faced with the heightened challenge of detecting the 
truly causative interactions while being confronted with the specious interactions involving only one of the marginal covariates. Such ``mixed" 
interactions (non-causative interactions involving one of $X_1$, $X_2$, $X_3$, or $X_4$; e.g. $X_1X_7$, $X_3X_9$, $X_2X_4$) can sometimes appear to be 
strongly associated with the response, when in fact the marginal effect of the causative covariate(s) alone is causing such an association. It should be 
noted that the marginal effects are constructed to be overall rather weak, and, consequently, will not be strongly detected by a marginal screening 
method. 
Once again, we report 
$\mathcal{P}_{j_1, j_2}$,  $\mathcal{P}_{\text{top}}$, and $\mathcal{P}_{\text{a}}$, as well as the 25\%, 50\%, 75\% and 90\% quantiles of the minimum 
number of interactions required to capture both causative interactions. These results are given in Table \ref{table:Sim5_P} (selection proportions) and 
\ref{table:Sim5_Q} (quantiles). Even with the addition of main effects, we see no drop in the overall accuracy of GenCorr in determining the true set of 
interactions. 
These results indicate that GenCorr (under either norm) is yet quite capable of detecting the causative interactions regardless of the addition of 
marginal effects.

\begin{table}[H]
\caption{Simulation 5 Selection Proportions}
\label{table:Sim5_P}
\vspace*{0.1cm}
\centering
\begin{tabular}{ |c c c c c c| } 
\hline
Model&Method&$\mathcal{P}_{1,2}$& $\mathcal{P}_{3,4}$  &$\mathcal{P}_{\text{top}}$ & $\mathcal{P}_{\text{a}}$ \\ 
 \hline\hline
 \multirow{ 2}{*}{5.A}&GenCorr-T&0.9600&0.9775 &0.9975 &0.9375 \\
 &GenCorr-F &0.9650  &0.9900  &0.9975 &0.9550 \\\hline
 \multirow{ 2}{*}{5.B}&GenCorr-T&0.8075 & 0.8025 & 0.9925 &0.6125 \\
 &GenCorr-F &0.8250  &0.8275 &0.9925 &0.6550 \\\hline
\multirow{ 2}{*}{5.C}&GenCorr-T &  0.8025 & 0.8300 & 0.9775 & 0.6350  \\
&GenCorr-F &0.8150  &0.8425 &0.9825 &0.6600 \\\hline 
\end{tabular}
\end{table}

\begin{table}[H]
\caption{Simulation 5 Quantiles}
\label{table:Sim5_Q}
\vspace*{0.1cm}
\centering
\begin{tabular}{ |c c c c c c| } 
\hline
Model&Method&$q_{25} $& $q_{50}$ & $q_{75}$& $q_{90}$ \\ 
 \hline\hline
\multirow{ 2}{*}{5.A}&GenCorr-T& 2.0&2.0&2.0&3.0\\
 &GenCorr-F & 2.0&2.0&2.0&3.0\\\hline
 \multirow{ 2}{*}{5.B}&GenCorr-T&2.0& 3.0& 16.0& 167.1 \\
 &GenCorr-F &2.0& 2.5 & 11.0 & 93.0\\\hline
\multirow{ 2}{*}{5.C}&GenCorr-T & 2.0& 3.0& 17.0& 137.9  \\
&GenCorr-F &2.0& 3.0&12.0&90.2 \\\hline 
\end{tabular}
\end{table}

Overall, each part of Simulation 5 demonstrates that GenCorr is an accurate and reliable method for detecting causative interaction for a multivaraite response. In each case, the Frobenius norm  produces results that slightly exceed in accuracy those produced when using the taxi-cab norm. 

\subsection{Conclusion on Simulations}
Throughout the various simulations presented above, GenCorr under the Frobenius matrix norm consistently obtains the best empirical results. The 
superiority of the Frobenius norm over the taxi-cab norm is likely due to the former's aforementioned ability to more strongly emphasize a covariate's 
high association with the response, while at the same time de-emphasizing those covariates which have a weak association with the components of the 
response. It is our suggestion that the Frobenius norm be favored in any standard implementations of GenCorr. 
\subsection{Real Data Analysis}

We now turn our attention to performing a real data analysis on data from a genome-wide association mapping of outbred NMRI mice.
This data set comes from the work  presented in \cite{Zhang2012Mice} and is available at \href{http://cgd.jax.org/datasets/phenotype/nmri.shtml}{http://cgd.jax.org/datasets/phenotype/nmri.shtml}. 
The response has seven components ($q = 7$), as outlined in Table \ref{table:MiceResponse}. 
These components are  observed on each of $n = 288$ individual mice and represent commonly measured phenotypic traits of mice. 
The values for SBP, DBP, and MAP are missing for two mice.
We used the \texttt{mice} package in R (an amusing naming coincidence for sure) under the default settings to impute these values for the mice for which 
they are missing. For a general reference on usage of the \texttt{mice} package, see \cite{JSSv045i03}. It should also be noted that we have omitted the 
ACR (urinary albumin-to-creatinine
ratio) values for each mouse, as 259 of the 288 observed mice (nearly 90\%) have ACR values equal to 0. The associated covariate space for the mice data consists of $p = 44,428$ SNPs to be examined for genetic association with the recorded phenotypic traits of the observed mice.   

 \begin{table}[H]
\caption{Observed Phenotypic Traits in NMRI Mice}
\label{table:MiceResponse}
\vspace*{0.1cm}
\centering
\begin{tabular}{ |c  c| } 
\hline
Trait Abbreviation& Description \\ 
 \hline\hline
 SBP & Systolic blood pressure \\
 DBP & Diastolic blood pressure \\
 MAP & Mean arterial pressure \\
 HDL & HDL (High-density lipoprotein) cholesterol \\
 CHL & Total cholesterol \\
 TRI & Triglyceride levels \\
 GLU & Glucose levels\\\hline
\end{tabular}
\end{table}  

Our empirical analysis will be broken down into two stages. In the first stage (\ref{1stStage}), we implement an iterative screening method employing 
GenCorr to find main effects. We will also apply GenCorr to screen for pairwise interactive effects between all $44,428$ SNPs. In the second stage 
(\ref{2ndStage}) of the analysis, we will outline a post-screening approach using penalized regression. 

\subsubsection{First Stage Analysis}\label{1stStage}
In the first stage of this analysis we will use GenCorr-F to screen for both marginal and interactive effects. To screen for marginal effects, we use the 
same general iterative approach of \cite{ZhongZhu2014}, where we replace their use of DC-SIS with GenCorr. This process is done as follows:
\begin{itemize}
\item Apply GenCorr to the full mouse data set. Let $d = 2[n/\log{n}] = 102$. We select $p_1 < d$ many predictors to act as our set of initial SNPs. As 
suggested in \cite{ZhongZhu2014}, the value of $p_1$ is a value strictly between $1$ and $d$ such that a linear regression model using the top (as 
determined by GenCorr) $p_1$-many SNPs has the minimal mean squared prediction error (MSPE). 
We determine such a value by iterating over each possible $p_1$ value ($1 < p_1 < d$), fitting a linear model of $\YY$ regressed on the top $p_1$ 
predictors (ordered by marginal utility score from GenCorr), and then recording the associated MSPE for each $p_1$. With each choice of $p_1$, we randomly 
select a training set consisting of 216 observations (75\% of the observed data) to fit the model. This model is then tested on a validation set of 72 
observations (25\% of the observed data). In our case, we found $p_1 = 85$.    

\vspace{0.5cm}

\item Denote by $X_1$ the $n \times p_1$ matrix formed by examining the observations of only the top $p_1$-many covariates and denote by $X_1^c$ the $n 
\times (p-p_1)$ matrix formed by the observations of the remaining $(p- p_1)$ covariates not found in the columns of $X_1$. 
Let \[X_{\text{new}} = \left(I_n - X_1(X_1^{\prime}X_1)^{\dagger}X_1^{\prime}\right)X_{1}^c,\]
where $A^{\dagger}$ indicates the Moore-Penrose pseudo inverse of a square matrix $A$. This means that $X_{\text{new}}$ will contain the residuals from 
regressing $X_{1}^c$ onto $X_{1}$. Apply GenCorr to $\YY$ and $X_{\text{new}}$, with $\YY$ still acting as the matrix of responses and $X_{\text{new}}$ 
acting as the matrix of predictor observations. Based on the scores obtained from GenCorr, select the top $(d- p_1)$-many SNPs, as ordered by this most 
recent run of GenCorr. 
\end{itemize}

We now have a total of $d = 102$ SNPs, selected using an iterative application of GenCorr. This completes the first stage of marginal feature selection. 
Our attention is next focused on the selection of pairwise interactions. Due to computational limitations, as well as lack of a theoretical basis, we do 
not use an iterative approach to select pairwise interactions. Instead, we will run the interactive version of GenCorr on the full mouse data and then 
directly select the top $d$-many interactions, as ordered by GenCorr. All told we have a set of 204 features (102 marginal, 102 interactive) from this 
first stage of the analysis. These 204 features will be further examined in stage two of the analysis, as given below.

\subsubsection{Second Stage Analysis}\label{2ndStage}
In this second analysis stage, we will take the 204 SNPs obtained in stage one and fit several elastic net \citep{elasticNet} and lasso \citep{Tibshirani} 
models to the data using the \texttt{glmnet} package in R. For further reference on this process see \cite{glmnetCite}. Three final models will be fit: 
elastic net with $\alpha = 0.4$, elastic net with $\alpha = 0.8$, and lasso ($\alpha = 1$). Here $\alpha$ is as given in the penalty function
\[\left((1-\alpha)/2\right)\lVert B \rVert_F^2 + \alpha\sum_{j = 1}^{2d}\lVert B_j \rVert_2,\]
where $B$ is the matrix of coefficients, $\lVert \cdot \rVert_F$ is the Frobenius matrix norm, and $\lVert B_j \rVert_2$ is a group-lasso penalty on each 
coefficient vector $B_j$ (the $j$th row vector of $B$) for a single predictor or a single pairwise interaction.

In each case, we will select the model associate with the $\lambda$ on the regularization path that minimizes the mean 10-fold cross validated error. The 
loss function used here for cross validation is the mean squared error (MSE). 
We will use a modified version of the corrected Akaike information criterion ($\text{AIC}_c$) to select what we determine to be the best candidate model (with smaller 
$\text{AIC}_c$ values being considered better). The $\text{AIC}_c$ has been shown to provide a more accurate estimation of model order than the standard 
AIC does ``when the number of fitted parameters is a moderate to large fraction of the sample size'' \citep{AICc1994}. The usual multivariate version of 
the $\text{AIC}_c$ is given as follows \citep{Seghouane2011new}:
\[{\rm AIC}_{c}=-2\ln\left(\mathcal{L}\right)+{2n(qk+q(q+1)/2)\over n-(k+q+1)},\]
where $\mathcal{L}$ is the maximum value of the likelihood function for the model in question. 
An equivalent (although unusual) definition of $\text{AIC}_c$ can be given as follows:
\[{\rm sAIC}_{c}=-2\left(\ln(\mathcal{L}) -\ln(\mathcal{L}_0)\right) +{2n(qk+q(q+1)/2)\over n-(k+q+1)},\]
where $\mathcal{L}_0$ denotes the maximum likelihood for the null model. We will call this version of the $\text{AIC}_c$ the \emph{shifted}-$\text{AIC}_c$ 
(abbreviated as $\text{sAIC}_c$). Note that for any candidate model, $\text{sAIC}_c = \text{AIC}_c + 2\ln(\mathcal{L}_0)$. 
Hence the ordering of candidate models given by the $\text{sAIC}_c$ will be equivalent to the ordering given by the usual definition of $\text{AIC}_c$, as 
the former is just a constant shift of the latter. We use the $\text{sAIC}_c$ for our model selection here, as it is easier to obtain from the results 
given in \texttt{glmnet}. The following lemma addresses these thoughts.
\begin{lemma}\label{saic}
Define the deviance of a fitted model by \[\mathcal{D} =2\left(\ln(\mathcal{L}_S) - \ln(\mathcal{L})\right), \]
where $\mathcal{L}_S$ is the maximum likelihood of the saturated model (the model with a free parameter for each observation). Define the null deviance to 
be
\[\mathcal{D}_0 =2\left(\ln(\mathcal{L}_S) - \ln(\mathcal{L}_0)\right).\]
Given $\mathcal{D}$ and $\mathcal{D}_0$, but not the log-likelihood of the fitted model directly, we can yet obtain the $\text{sAIC}_c$. 
\begin{proof}
Note that $\mathcal{D}_0 - \mathcal{D} = 2\left(\ln(\mathcal{L}) -\ln(\mathcal{L}_0)\right).$ Thus we now can determine the $\text{sAIC}_c$ as below:
\[{\rm sAIC}_{c}=-2\left(\ln(\mathcal{L}) -\ln(\mathcal{L}_0)\right) +{2n(qk+q(q+1)/2)\over n-(k+q+1)}.\]
\end{proof}
\end{lemma}
Although the \texttt{glmnet} package in R does not provide us with direct access to the log-likelihood of any candidate model, the package does allow us 
to obtain both $\mathcal{D}$ and $\mathcal{D}_0$ explicitly. By Lemma \ref{saic}, we can thus find the associated $\text{sAIC}_c$ for any candidate model. 
To alleviate the issue of having to compare $\text{sAIC}_c$ values whose lower order (ones, tens, hundreds) place-values do not make a substantive 
difference in determining model ordering, we will divide each raw $\text{sAIC}_c$ value by 1000 when reporting final results. This scaling of course does 
not affect the overall ordering of candidate models and makes the results easier to visually parse. 

\subsubsection{Results of Real Data Analysis.} We report the final outcome of our two stage process below in Table \ref{table:RealData}.    
We denote the three different models by their method of second stage analysis. (The first stage was the same for each model and was only performed once). 
As measures of model fitness, we report the mean 10-fold cross validated MSE and the (scaled, see above) $\text{sAIC}_c$. We also report for each 
component of the response the number of features (marginal, interactive, total) retained by the individual penalized regression approaches.

\setlength{\tabcolsep}{4.25pt}
\begin{table}[H]
\caption{Results of Real Data Analysis}
\label{table:RealData}
\vspace*{0.1cm}
\centering
\begin{tabular}{ |c |r c c c c c c c| } 
\hline
\textit{Second Stage}&\textit{Features}&\textit{SBP}&\textit{DBP}&\textit{MAP}&\textit{HDL}&\textit{CHL}&\textit{TRI}&\textit{GLU}\\ 
 \hline\hline
\textbf{El. Net} ($\bm{\alpha = 0.4}$)&Marginal:&102&102&102&102&102&102&102\\
MSE $= 5873.393$&Interact:&82&82&82&79&79&81&82\\
$\text{sAIC}_c = -1558.766$ &Total:&184&184&184&181&181&183&184\\\hline
 \textbf{El. Net} ($\bm{\alpha = 0.8}$)&Marginal:&43&43&43&43&43&43&43\\
 MSE $= 5989.832 $&Interact:&16&16&16&16&16&17&17\\
$\text{sAIC}_c = -1565.601$ &Total:&59&59&59&59&59&60&60\\\hline
 \textbf{Lasso} &Marginal:&29&29&29&29&29&29&29\\
 MSE $= 5696.422$ &Interact:&4&4&4&4&4&4&4\\
$\text{sAIC}_c = -1566.418$ &Total:&33&33&33&33&33&33&33\\ \hline 
\end{tabular}
\end{table}

\setlength{\tabcolsep}{6pt}

Except for the few cases in the two elastic net models where the number of retained interactions differs slightly across the components of the response, the same features were retained for each of the seven response variables. In the cases where some components were associated with slightly more non-zero coefficients than the other components, these ``extra'' non-zero coefficients were the only places where the selected features differ across component.  
When we used only a moderate mixing parameter in elastic net $(\alpha = 0.4)$, few of the features are dropped from the model (and the only dropped 
features are interactions). 
This gave us around an 11\% reduction in total model size. Because the aim of penalized regularization regression is to obtain a more parsimonious 
model, the failure to substantively reduce the model size is undesirable. When we increased the mixing parameter to $\alpha = 0.8$, we were able to obtain 
a more favorable reduction in total model size. 
However, while such reduction in size was desirable (see for example the lower $\text{sAIC}_c$ for this model 
compared to the previous model), it also came at the cost of increased mean cross validated MSE, 
leading us to be hesitant about fully embracing this model. The candidate model that we deemed to be the most advisable overall is that which was obtained
by using lasso in the second stage of the analysis. This model possess the lowest $\text{sAIC}_c$ value, as well as the lowest mean cross validated MSE. 
Moreover, this model accomplished by far the largest reduction in model size out of the the three approaches, 
yielding a model that is not only superior in terms of mean cross validated MSE and $\text{sAIC}_c$, but also salient in its parsimony.  

\section{Discussion}\label{discussion}
In this paper we proposed a new feature screening approach, called GenCorr, which is applicable to ultrahigh dimensional data with multivariate response. 
Our method allows us to perform both marginal and interactive screening all within the same overall methodological framework. We have demonstrated the 
finite performance of GenCorr under a series of empirical simulations. In the marginal case, we compared our results with those of SIRS 
\citep{ZhuLiLiZhu2011} and DC-SIS \citep{DCSIScitation}. We also presented a real-data analysis, showing the application of GenCorr to data originating 
from a GWAS setting.  From a theoretical perspective, we have shown that GenCorr possess the strong sure screening property, that is, with probability 
converging to one asymptotically, GenCorr selects the true model exactly. 

In this paper we have avoided directly selecting a cutoff for model selection. Several approaches have been proposed for selecting such a cutoff: 
\cite{ZhuLiLiZhu2011} submitted one possible method for choosing a cutoff for SIRS; \cite{Huang2014} developed another approach for choosing a selection 
cutoff for their Pearson chi-squared-test-based screening method; 
\cite{KongWangWahba2015} present an iterative approach to producing an implicit selection cutoff. One may explore these cutoff methods further and adapt 
them as they see fit. We will not further pursue the topic here, however.

\section{Proofs of Theorems \ref{thm1} and \ref{thm2}}\label{proofs}
Here we present the proofs of Theorems \ref{thm1} and \ref{thm2} as presented in Section \ref{methods}. All proofs here will be presented in the most 
general form of the context of screening for an $r$-way interaction between covariates $X_{j_1}$, $X_{j_2},\ldots,$ $X_{j_r}$. When $r=1$, we have the 
necessary results for marginal effects screening. Throughout this section, $\pnorm{\cdot}$ will refer to any $\mathfrak{p}$-norm with $\mathfrak{p}$ 
finite. For any positive integer $k$, $\pnorm{\cdot}$ can be viewed as a function from $\RR^k$ to $\RR$. This means that $(\RR^k, \pnorm{\cdot})$ forms a 
normed vector space. This leads us to a routine lemma. 

\subsection{Prefacing Lemmas}

\begin{lemma}\label{lem1}
Let $k$ be any positive integer. The $\mathfrak{p}$-norm $\pnorm{\cdot}$ is a continuous function from $\RR^k$ to $\RR$. 
\begin{proof}
Take any $\varepsilon > 0$. Suppose that $\{\bm{a}_N\}_{N = 1}^{\infty}$ is a sequence in $\RR^k$ with $\lim_{N \to \infty}\bm{a}_N = \bm{a}$. This means 
that there exists some positive integer $N_0$ such that whenever $N > N_0$, we have that \[\pnorm{\bm{a}_N - \bm{a}} < \varepsilon.\] However, by the 
reverse triangle inequality, we know that \[\left|\pnorm{\bm{a}_N} - \pnorm{\bm{a}}\right| \leq \pnorm{\bm{a}_N - \bm{a}}\] for any positive integer $N$. 
Thus there exists some positive integer $N_0$ (the same one as determined above in fact) such that whenever $N > N_0$, we have
\[\left|\pnorm{\bm{a}_N} - \pnorm{\bm{a}}\right| < \varepsilon.\] Thus $\lim_{N \to \infty} \pnorm{\bm{a}_N} = \pnorm{\bm{a}}$, completing the proof. 
\end{proof}
\end{lemma}

At multiple times throughout this section we will employ the continuous mapping theorem. For further reference on the continuous mapping theorem, see e.g. 
\cite{Serfling1980} and \cite{casella2002statistical}. 
The well known weak law of large numbers will also be used several times to establish the consistency of various sample estimators. Below, we provide a 
slightly modified version of the weak law of large numbers. 

\begin{lemma}\label{wlln}
Given an independent and identically distributed collection $\{W_1, W_2, \ldots, W_n\}$ of $n$-many samples of a random variable $W$ with $\EE W = \mu$ 
and $\text{Var}(W) = \sigma^2 < \infty$, define \[\widetilde{W} = \frac{1}{n-1}\sum_{i = 1}^n W_i.\] We then have that $\widetilde{W} \xrightarrow{P} \EE 
W$ as $n \to \infty$.
\begin{proof}
Because the samples of $W$ are all identically distributed, we have know that $\text{Var}(W_i) = \sigma^2$ for all $i$. Let $\overline{W} = \frac{1}
{n}\sum_{i = 1}^n W_i$ be the usual sample mean. Due to the independence of $W_1$, $W_2, \ldots, W_n$, we have the following:
\begin{align*}
\text{Var}\left(\overline{W}\right) &=\text{Var}\left(\frac{1}{n}(W_1 + W_2 + \cdots + W_n)\right)\\[0.75ex]
&= \frac{1}{n^2}\text{Var}\left(W_1 + W_2 + \cdots + W_n\right) \\[0.75ex]
&= \frac{n\sigma^2}{n^2}\\[0.75ex]
&= \frac{\sigma^2}{n}.
\end{align*}
The common mean of each $W_i$ is the same as the expected value of $\overline{W}$, namely $\EE\overline{W} = \mu$. By Chebyshev's inequality and for any 
$\varepsilon > 0$, we have
\[\PP\left(|\overline{W} - \mu|\geq \varepsilon\right) \leq \frac{\sigma^2}{n\varepsilon^2}.\]
This in turn implies the following:
\begin{equation}\label{oneminus}
\PP\left(|\overline{W} - \mu|< \varepsilon\right) = 1-\PP\left(|\overline{W} - \mu|\geq \varepsilon\right) \geq 1- \frac{\sigma^2}{n\varepsilon^2}.
\end{equation}
Taking the limit as $n\to \infty$ in (\ref{oneminus}), we obtain
\[
\lim_{n \to \infty}\PP\left(|\overline{W} - \mu|< \varepsilon\right) \geq \lim_{n \to \infty}\left(1- \frac{\sigma^2}{n\varepsilon^2}\right)
= 1 - \lim_{n \to \infty}\frac{\sigma^2}{n\varepsilon^2} = 1.
\]
Thus $\overline{W} \xrightarrow{P} \mu$. Note that $\widetilde{W} = \frac{n}{n-1}\overline{W}$. It is easily seen that the numeric sequence $\{\frac{n}
{n-1}\}_{n = 1}^{\infty}$ converges to one as $n$ approaches infinity. By a standard application of the continuous mapping theorem, we know the following:
\[\widetilde{W}  = \frac{n}{n-1}\overline{W} \xrightarrow{P} 1\cdot \mu = \mu.\]
In conclusion, we have $\widetilde{W} \xrightarrow{P} \mu$, which is the desired result. 
\end{proof} 
\end{lemma}

We remind the reader of the matrix $\Sigma_{j_1, \ldots, j_r}$ as previously defined in Section \ref{methods}.
The entries of $\Sigma_{j_1, \ldots, j_r}$ are of three main forms: 
\begin{itemize}
\item The covariance between two components of $\YY$. As given in Section \ref{methods}, this covariance is estimated by the sample covariance given as follows:
\begin{equation}\label{covar}
\widehat{\text{Cov}}\left(Y^{(\ell)}, Y^{(m)}\right) = \frac{1}{n-1} \sum_{i = 1}^n \left(Y^{(\ell)}_i - \overline{Y}^{(\ell)}\right)\left(Y^{(m)}_i - \overline{Y}^{(m)}\right),
\end{equation}
where $\overline{Y}^{(\ell)}$ and $\overline{Y}^{(m)}$ are the sample means of $Y^{(\ell)}$ and $Y^{(m)}$ respectively. 

\item The product of the variances of each of $X_{j_1},\ldots,$ $X_{j_r}$.
As given in Section \ref{methods}, each variance is estimated by the sample variance given below:
\begin{equation}\label{varprod}
\widehat{\text{Var}}\left(X_{j_s}\right) = \frac{1}{n-1} \sum_{i = 1}^n \left(X_{ij_s} - \overline{X}_{j_s}\right)^2,
\end{equation}
where $\overline{X}_{j_s}$ is the standard sample mean of $X_{j_s}$.

\item The $(r+1)$-way joint cumulant between a component of $\YY$ and the random variables $X_{j_1},\ldots,$ $X_{j_r}$. The $({r+1})$-way joint cumulant can be estimated as follows:
\begin{equation}\label{jointcum}
\small\widehat{\kappa}_{r+1}(Y^{(m)}, X_{j_1}, \ldots, X_{j_r}) = \frac{1}{n} \sum_{i = 1}^n \left[~\prod_{s=1}^r\left(X_{ij_s} - \overline{X}_{j_s}\right)\right]\left(Y^{(m)}_i - \overline{Y}^{(m)}\right),
\end{equation}
\normalsize
where $\overline{X}_{j_s}$ represents the sample mean of $X_{j_s}$ and $\overline{Y}^{(m)}$ is the sample mean of $Y^{(m)}$.
\end{itemize}

It is an overall straightforward exercise to verify that each of (\ref{covar}), (\ref{varprod}), and (\ref{jointcum}) are consistent estimators of their associated population parameters. We present this in the form of a three-part lemma. 

\begin{lemma}\label{consistency}
With the components of $\YY$ being given by $Y^{(m)}$ ($m = 1,2,\ldots, q$), and each covariate in question being denoted by $X_{j_s}$ ($s = 1, 2, \ldots, r$), we have the following results:
\begin{enumerate}
\item $\widehat{\text{Cov}}\left(Y^{(\ell)}, Y^{(m)}\right) \xrightarrow{P} {\text{Cov}}\left(Y^{(\ell)}, Y^{(m)}\right)$;\vspace{0.333cm}
\item $\prod_{s = 1}^r\widehat{\text{Var}}\left(X_{j_s}\right) \xrightarrow{P} \prod_{s = 1}^r {\text{Var}}\left(X_{j_s}\right)$;\vspace{0.333cm}
\item $\widehat{\kappa}_{r+1}(Y^{(m)}, X_{j_1}, \ldots, X_{j_r})\xrightarrow{P} {\kappa}_{r+1}(Y^{(m)}, X_{j_1}, \ldots, X_{j_r}).$ 
\end{enumerate}
\begin{proof}
Statement 1 can be demonstrated as follows: Expanding the right-hand side of (\ref{covar}), we have 
\small
\begin{align}
\begin{split}\label{expanded}
\widehat{\text{Cov}}\left(Y^{(\ell)}, Y^{(m)}\right) 
&= \frac{1}{n-1}\sum_{i = 1}^n Y^{(\ell)}_i Y^{(m)}_i - \frac{1}{n-1}\sum_{i = 1}^nY^{(\ell)}_i\overline{Y}^{(m)}\\
   &\quad \quad \quad\quad - \frac{1}{n-1}\sum_{i = 1}^n \overline{Y}^{(\ell)} Y^{(m)}_i + \frac{1}{n-1}\sum_{i = 1}^n \overline{Y}^{(\ell)} \overline{Y}^{(m)}
   \end{split}
\end{align}
\normalsize
By applying the weak law of large numbers as presented in Lemma \ref{wlln} term-wise to the right hand side of (\ref{expanded}),  we obtain
\begin{align*}
\frac{1}{n-1}\sum_{i = 1}^n Y^{(\ell)}_i Y^{(m)}_i &\xrightarrow{P} \EE\left(Y^{(\ell)}Y^{(m)}\right)\\[1.3ex]
\frac{1}{n-1}\sum_{i = 1}^nY^{(\ell)}_i\overline{Y}^{(m)}&\xrightarrow{P} \EE\left(Y^{(\ell)}\right)\EE\left(Y^{(m)}\right)\\[1.3ex]
\frac{1}{n-1}\sum_{i = 1}^n \overline{Y}^{(\ell)} Y^{(m)}_i&\xrightarrow{P} \EE\left(Y^{(\ell)}\right)\EE\left(Y^{(m)}\right)\\[1.3ex]
\frac{1}{n-1}\sum_{i = 1}^n \overline{Y}^{(\ell)} \overline{Y}^{(m)}&\xrightarrow{P} \EE\left(Y^{(\ell)}\right)\EE\left(Y^{(m)}\right).
\end{align*}
By the continuous mapping theorem applied to addition and subtraction of estimators, we now have

\small
\begin{align*}\widehat{\text{Cov}}\left(Y^{(\ell)}, Y^{(m)}\right) &\xrightarrow{P} \left[\EE\left(Y^{(\ell)}Y^{(m)}\right) - 2\EE\left(Y^{(\ell)}\right)\EE\left(Y^{(m)}\right) + \EE\left(Y^{(\ell)}\right)\EE\left(Y^{(m)}\right)\right]\\[1.3ex]
&=\EE\left(Y^{(\ell)}Y^{(m)}\right) - \EE\left(Y^{(\ell)}\right)\EE\left(Y^{(m)}\right)\\[1.3ex]
&= {\text{Cov}}\left(Y^{(\ell)}, Y^{(m)}\right).
\end{align*}

\normalsize

Statement 2 is a direct corollary of statement 1. Each multiplicand in the product $\prod_{s = 1}^r\widehat{\text{Var}}\left(X_{j_s}\right)$ can be shown 
to converge in probability to the associated multiplicand in $\prod_{s = 1}^r{\text{Var}}\left(X_{j_s}\right)$. This is done by viewing 
$\text{Var}\left(X_{j_s}\right)$ as the covariance between $X_{j_s}$ and itself,
then proceeding quite similarly as was done in proving statement 1 of the lemma. This 
gives us that $\widehat{\text{Var}}\left(X_{j_s}\right) \xrightarrow{P} {\text{Var}}\left(X_{j_s}\right)$ for each $s = 1,2,\ldots, r$. By a simple 
application of the continuous mapping theorem, the product of these estimators of the variances converges in probability to the product $\prod_{s = 
1}^r{\text{Var}}\left(X_{j_s}\right)$. 

Statement 3 can be established using a similar approach to that of statement 1. By expanding the product in the right hand side of (\ref{jointcum}) we obtain terms with one of the two following general forms, with the covariates $X_{j_1}, X_{j_2}, \ldots, X_{j_s}$ being (without loss of generality) ordered here to simplify the notation:
\begin{equation}\label{type1}
(-1)^{r-s}~\frac{1}{n}\sum_{i = 1}^n X_{ij_1}X_{ij_2}\cdots X_{ij_s}\overline{X}_{j_{(s+1)}}\overline{X}_{j_{(s+2)}}\cdots\overline{X}_{j_{r}}Y^{(m)}_i
\end{equation} 
or
\begin{equation}\label{type2}
(-1)^{r-s-1}~\frac{1}{n}\sum_{i = 1}^n X_{ij_1}X_{ij_2}\cdots X_{ij_s}\overline{X}_{j_{(s+1)}}\overline{X}_{j_{(s+2)}}\cdots\overline{X}_{j_{r}}\overline{Y}^{(m)}.
\end{equation} 
The terms (\ref{type1}) and (\ref{type2}) represent the general form of each summand of the right hand side of (\ref{jointcum}) when expanded out fully. By a standard application of the (more traditional) weak law of large numbers, we have the following:

\small
\begin{equation*}
\begin{split}
(\ref{type1}) \xrightarrow{P} (-1)^{r-s}~\EE\left(X_{j_1}X_{j_2}\cdots X_{j_s}Y^{(m)}\right)\EE\left({X}_{j_{(s+1)}}\right)\EE\left({X}_{j_{(s+2)}}\right)\cdots\EE\left({X}_{j_{r}}\right);\\[1.3ex]
%
(\ref{type2}) \xrightarrow{P} (-1)^{r-s-1}~ \EE\left(X_{j_1}X_{j_2}\cdots
X_{j_s}\right)\EE\left({X}_{j_{(s+1)}}\right)\EE\left({X}_{j_{(s+2)}}\right)\cdots\EE\left({X}_{j_{r}}\right)\EE\left(Y^{(m)}\right). 
\end{split}\end{equation*}

\normalsize
Remembering that \[{\kappa}_{r+1}(Y^{(m)}, X_{j_1}, \ldots, X_{j_r})  = \EE\left(\left(Y^{(m)} - \EE{Y}^{(m)}\right)\prod_{s = 1}^r \left(X_{j_s} - \EE{X}_{j_s}\right)\right)\] and combining the above statements on the convergence of (\ref{type1}) and (\ref{type2}), then applying the continuous mapping theorem, we finally obtain
\[\widehat{\kappa}_{r+1}(Y^{(m)}, X_{j_1}, \ldots, X_{j_r})\xrightarrow{P} {\kappa}_{r+1}(Y^{(m)}, X_{j_1}, \ldots, X_{j_r}).\]
This completes the proof of the three statements of the lemma.
\end{proof}
\end{lemma}

\begin{lemma}\label{PhihatConsistant}
As an estimator of $\Phi_{j_1, \ldots, j_r}$, $\PhiHat$ is consistent. 
\begin{proof}
Lemma \ref{consistency} establishes that every entry of $\widehat{\Sigma}_{j_1,  \ldots, j_r}$ is a consistent estimator of the associated entry of $\Sigma_{j_1, \ldots, j_r}$. By the continuous mapping theorem, this means that every entry in the matrix \[\widehat{\mathcal{H}}_{j_1, \ldots, j_r} = \left[\text{diag}\left(\widehat{\Sigma}_{j_1, \ldots, j_r}\right)\right]^{-1/2} \widehat{\Sigma}_{j_1, \ldots, j_r}\left[\text{diag}\left(\widehat{\Sigma}_{j_1, \ldots, j_r}\right)\right]^{-1/2}\]
as defined previously in Section \ref{methods} is a consistent estimators of the entries of 
\[\mathcal{H}_{j_1, \ldots, j_r} = \left[\text{diag}\left({\Sigma}_{j_1, \ldots, j_r}\right)\right]^{-1/2} {\Sigma}_{j_1, \ldots, j_r}\left[\text{diag}\left({\Sigma}_{j_1, \ldots, j_r}\right)\right]^{-1/2}.\] 
This can be seen by first noting that products and sums of consistent estimators are also consistent estimators themselves and that the diagonal entries of $\Sigma_{j_1, \ldots, j_r}$ are all positive. As the functions $f(t) = 1/t$ and $g(t) = \sqrt{t}$ are both continuous for positive values of $t$, the continuous mapping theorem indeed ultimately gives the results on the consistency of the entries of $\widehat{\mathcal{H}}_{j_1, \ldots, j_r}$ as estimators of the respective entries of $\mathcal{H}_{j_1, \ldots, j_r}$. 

Note that for our specific application here, $\lVert \cdot \rVert_{\mathfrak{p}}$ is a function of $(q+1)^2$-many variables. This comes from the fact that $\mathcal{H}_{j_1, \ldots, j_r}$ is a $(q+1) \times (q+1)$ dimensional matrix. 
As such, by Lemma \ref{lem1} we can apply $\lVert\cdot\rVert_{\mathfrak{p}}$ to the matrix $\widehat{\mathcal{H}}_{j_1, \ldots, j_r}$ and obtain (again by continuous mapping theorem) that $\PhiHat$ is a consistent estimator of 
$\Phi_{j_1, \ldots, j_r}$. 
\end{proof}
\end{lemma}

\subsection{Proofs of Main Results}
Before proceeding further, we remind the reader of the definition of the following constant:
\[\gamma = (q+1) + \sum_{1\leq \ell, m \leq q} |\rho_{\ell m}|^2,\] where $\rho_{\ell m}$ is the correlation between $Y^{(\ell)}$ and $Y^{(m)}$.
 This constant will be referenced several times throughout the proofs of Theorems \ref{thm1} and \ref{thm2}. 
 
 The proofs of the main results will be presented in four steps.
 \begin{itemize}
\item \textsc{Step 1}: We will show that there exists a positive constant $\Phi_{\min}$ such that for any tuple $(j_1, \ldots, j_r) \in \mathcal{S}_T$,
\[\PhiPop > \Phi_{\min} > \sqrt{\gamma} > 0.\]
 (Note that this is also Corollary \ref{cor1}).  
 In Condition (C2), we defined the value $\omega_{j_1, \ldots, j_r} > 0$ such that for some component, $Y^{(m)}$, of $\YY$, we have  
 \[\left|\frac{\kappa_{r+1}\left(Y^{(m)}, X_{j_1}, \ldots, X_{j_r}\right)}{\sigma_{(m)}\sigma_{j_1}\cdots\sigma_{j_r}}\right| > \omega_{j_1, \ldots, j_r}>0,\]
 whenever $(j_1,\ldots, j_r)$ is in $\mathcal{S}_T$.
As an aside, note that  $|\mathcal{S}_T| \leq \binom{p}{r} < \infty$, as there can only be $\binom{p}{r}$ many $r$-way interactions formed among $p$-many covariates.
  Define a value $\omega_{\min}$ as follows:
 \[\omega_{\min} = \min_{(j_1, \ldots, j_r) \in \mathcal{S}_t}\left\{\omega_{j_1, \ldots, j_r}\right\}.\]
Because each $\omega_{j_1, \ldots, j_r}$ is positive and because $|\mathcal{S}_T| \leq \infty,$ the minimum is a well defined and positive value here. In 
other words, we have some $\omega_{\min} > 0$ such that $\omega_{j_1, \ldots, j_r} > \omega_{\min}$ for all $(j_1, \ldots, j_r) \in \mathcal{S}_T$. It 
therefore can be seen that for any $(j_1, \ldots, j_r) \in \mathcal{S}_T$ we have the following chain of inequalities:
\begin{align*}
\PhiPop & = \pnorm{\mathcal{H}_{j_1, \ldots, j_r}}\\[0.65ex]
&\geq \sqrt{(\omega_{j_1, \ldots, j_r})^2 + \gamma}\\[0.65ex]
&> \sqrt{\omega_{\min}^2/2 + \gamma}\\[0.65ex]
&> \sqrt{\gamma}.
\end{align*} 

Let $\Phi_{\min} =   \sqrt{\omega_{\min}^2/2 + \gamma}$. Then, for all $r$-tuples in the true model,
\[\PhiPop > \Phi_{\min} > \sqrt{\gamma} > 0.\]
This completes Step 1, as well as proves Corollary \ref{cor1}.  
 
\vspace*{0.5cm} 
 
 \item \textsc{Step 2}: 
 In line with the statement of Corollary \ref{cor2}, we will now show that $\PhiHat$ is a uniformly consistent estimator of $\PhiPop$. Lemma \ref{consistency} tells us that $\PhiHat$ is a consistent estimator of $\PhiPop$. Thus for any $(j_1, \ldots, j_r) \in \mathcal{I}$ and any $\varepsilon > 0$, we know
 \[\lim_{n \to \infty} \PP\left(\left|\PhiHat - \Phi_{j_1, \ldots, j_r}\right| > \varepsilon \right) = 0.\]
Let $(J_1, \ldots, J_r)$ be the $r$-tuple in $\mathcal{I}$ that maximizes $\left|\PhiHat - \Phi_{j_1, \ldots, j_r}\right|$. By Lemma \ref{consistency}, we know that
\[\lim_{n \to \infty} \PP\left(\left|\PhiHat - \Phi_{J_1, \ldots, J_r}\right| > \varepsilon \right) = 0.\]
Thus, for any $\varepsilon > 0$, we have
\[\lim_{n \to \infty} \PP\left(\max_{(j_1,\ldots, j_r) \in \mathcal{I}}\left|\PhiHat - \Phi_{j_1, \ldots, j_r}\right| > \varepsilon \right) = 0.\] This establishes that $\PhiHat$ is a uniformly consistent estimator of $\PhiPop$. This completes Step 2, as well as the proof of Corollary \ref{cor2}. 
 
\vspace*{0.5cm} 
 
 \item \textsc{Step 3}: We now show that there is a positive constant $c$ such that 
 \[\lim_{n \to \infty} \PP\left(\mathcal{S}_T \subseteq \widehat{\mathcal{S}}\right) = 1.\]
 Let $c = \Phi_{\min}$. By way of contradiction, suppose that this selection of $c$ does not result in $\mathcal{S}_T \subseteq \widehat{S}$. This means that we can find some $(j_1^*, \ldots, j_r^*) \in \mathcal{S}_T$, while at the same time $(j_1^*, \ldots, j_r^*) \not\in \widehat{\mathcal{S}}$. By the definition of $\widehat{\mathcal{S}}$, we then know that \[\widehat{\Phi}_{j_1^*, \ldots, j_r^*} \leq \Phi_{min},\]
 and also \[\Phi_{j_1^*, \ldots, j_r^*} > \Phi_{\min}.\]
It now follows that \[\max_{(j_1,\ldots, j_r) \in \mathcal{I}}\left|\PhiHat - \Phi_{j_1, \ldots, j_r}\right| \geq \left|\widehat{\Phi}_{j_1^*, \ldots, j_r^*} - \Phi_{j_1^*, \ldots, j_r^*}\right| > \Phi_{\min}. \]
However, by uniform consistency shown in Step 2,
\small
\[\PP\left(\mathcal{S}_T \not\subseteq \widehat{\mathcal{S}}\right) \leq \PP\left(\max_{(j_1,\ldots, j_r) \in \mathcal{I}}\left|\PhiHat - \Phi_{j_1, \ldots, j_r}\right| > {\Phi_{\min}} \right) \to 0,  \text{ as } n \to \infty.\] 
\normalsize 
This contradicts are previous supposition of non-containment. Thus we have 
\[\lim_{n \to \infty}\PP\left(\mathcal{S}_T \subseteq \widehat{\mathcal{S}}\right) = 1,\]
finishing the proof to Theorem \ref{thm1} and showing the forward direction to Theorem \ref{thm2}. 

\vspace*{0.5cm} 

\item \textsc{Step 4:} We now show the reverse direction for Theorem \ref{thm2}. Suppose once again by way of contradiction that $\widehat{\mathcal{S}} \not\subseteq \mathcal{S}_T$. Then there exists an $r$-tuple $(j_1^*, \ldots, j_r^*) \in \widehat{\mathcal{S}}$, yet $(j_1^*, \ldots, j_r^*) \not\in \mathcal{S}_T$. It follows that 
\[\widehat{\Phi}_{j_1^*, \ldots, j_r^*} > \Phi_{\min} > \sqrt{\gamma}.\] However, by Condition (C3), we assumed that $\Phi_{j_1^*, \ldots, j_r^*} = \sqrt{\gamma}$.  
 Let $\varepsilon = \left(\Phi_{\min} - \sqrt{\gamma}\right)/2$. This value of $\varepsilon$ is verifiably greater than zero. Note that 
 \[\max_{(j_1,\ldots, j_r) \in \mathcal{I}}\left|\PhiHat - \Phi_{j_1, \ldots, j_r}\right| \geq \left|\widehat{\Phi}_{j_1^*, \ldots, j_r^*} - \Phi_{j_1^*, \ldots, j_r^*}\right| > \varepsilon.\]
 However, by uniform consistency, we also have 
\small 
 \[\PP\left(\widehat{\mathcal{S}} \not\subseteq \mathcal{S}_T\right) \leq \PP\left(\max_{(j_1,\ldots, j_r) \in \mathcal{I}}\left|\PhiHat - \Phi_{j_1, \ldots, j_r}\right|> \varepsilon \right) \to 0, \text{ as } n \to \infty.\]
 \normalsize
 This causes a contradiction with our assumption of non-containment and $\max_{(j_1,\ldots, j_r) \in \mathcal{I}}\left|\PhiHat - \Phi_{j_1, \ldots, j_r}\right| > \varepsilon$. Therefore, we can in fact conclude that 
 \[\lim_{n \to \infty}\PP\left(\widehat{\mathcal{S}}\subseteq \mathcal{S}_T\right) = 1.\]
 
 The combination of Steps 3 and 4 shows that the selection of $c = \Phi_{\min}$ yields the full result of Theorem \ref{thm2}, namely
 \[\lim_{n \to \infty}\PP\left(\mathcal{S}_T = \widehat{\mathcal{S}}\right) = 1.\]
 This completes the requisite proofs.  
  
\end{itemize} 

\bibliographystyle{imsart-nameyear}

\bibliography{BibliographyGC}

\begin{thebibliography}{32}

\bibitem[\protect\citeauthoryear{Bedrick and Tsai}{1994}]{AICc1994}
\begin{barticle}[author]
\bauthor{\bsnm{Bedrick},~\bfnm{Edward~J.}\binits{E.~J.}} \AND
  \bauthor{\bsnm{Tsai},~\bfnm{Chih-Ling}\binits{C.-L.}}
(\byear{1994}).
\btitle{Model Selection for Multivariate Regression in Small Samples}.
\bjournal{Biometrics}
\bvolume{50}
\bpages{226--231}.
\end{barticle}
\endbibitem

\bibitem[\protect\citeauthoryear{Casella and
  Berger}{2002}]{casella2002statistical}
\begin{bbook}[author]
\bauthor{\bsnm{Casella},~\bfnm{G.}\binits{G.}} \AND
  \bauthor{\bsnm{Berger},~\bfnm{R.~L.}\binits{R.~L.}}
(\byear{2002}).
\btitle{Statistical Inference}.
\bseries{Duxbury advanced series in statistics and decision sciences}.
\bpublisher{Thomson Learning}.
\end{bbook}
\endbibitem

\bibitem[\protect\citeauthoryear{Chu, Li and Reimherr}{2016}]{Chu2016feature}
\begin{barticle}[author]
\bauthor{\bsnm{Chu},~\bfnm{Wanghuan}\binits{W.}},
  \bauthor{\bsnm{Li},~\bfnm{Runze}\binits{R.}} \AND
  \bauthor{\bsnm{Reimherr},~\bfnm{Matthew}\binits{M.}}
(\byear{2016}).
\btitle{Feature screening for time-varying coefficient models with ultrahigh
  dimensional longitudinal data}.
\bjournal{The annals of applied statistics}
\bvolume{10}
\bpages{596}.
\end{barticle}
\endbibitem

\bibitem[\protect\citeauthoryear{Cui, Li and Zhong}{2015}]{CuiLiZhong2015}
\begin{barticle}[author]
\bauthor{\bsnm{Cui},~\bfnm{Hengjian}\binits{H.}},
  \bauthor{\bsnm{Li},~\bfnm{Runze}\binits{R.}} \AND
  \bauthor{\bsnm{Zhong},~\bfnm{Wei}\binits{W.}}
(\byear{2015}).
\btitle{Model-Free Feature Screening for Ultrahigh Dimensional Discriminant
  Analysis}.
\bjournal{Journal of the American Statistical Association}
\bvolume{110}
\bpages{630-641}.
\end{barticle}
\endbibitem

\bibitem[\protect\citeauthoryear{Fan, Han and Liu}{2014}]{FanHanLiu:BigData}
\begin{barticle}[author]
\bauthor{\bsnm{Fan},~\bfnm{J.}\binits{J.}},
  \bauthor{\bsnm{Han},~\bfnm{F.}\binits{F.}} \AND
  \bauthor{\bsnm{Liu},~\bfnm{H.}\binits{H.}}
(\byear{2014}).
\btitle{Challenges of big data analysis}.
\bjournal{National Science Review}
\bvolume{1}
\bpages{293-314}.
\end{barticle}
\endbibitem

\bibitem[\protect\citeauthoryear{Fan and Li}{2001}]{SCAD2001}
\begin{barticle}[author]
\bauthor{\bsnm{Fan},~\bfnm{J.}\binits{J.}} \AND
  \bauthor{\bsnm{Li},~\bfnm{R.}\binits{R.}}
(\byear{2001}).
\btitle{Variable selection via nonconcave penalized likelihood and its oracle
  properties.}
\bjournal{Journal of the American Statistical Association}
\bvolume{96}
\bpages{1348-1360}.
\end{barticle}
\endbibitem

\bibitem[\protect\citeauthoryear{Fan and Lv}{2008}]{FanLv:2008}
\begin{barticle}[author]
\bauthor{\bsnm{Fan},~\bfnm{J.}\binits{J.}} \AND
  \bauthor{\bsnm{Lv},~\bfnm{J.}\binits{J.}}
(\byear{2008}).
\btitle{Sure independence screening for ultrahigh dimensional feature space}.
\bjournal{Journal of the Royal Statistical Society, Series B}
\bvolume{70}
\bpages{849-911}.
\end{barticle}
\endbibitem

\bibitem[\protect\citeauthoryear{Fan and Lv}{2010}]{fan2010selective}
\begin{barticle}[author]
\bauthor{\bsnm{Fan},~\bfnm{Jianqing}\binits{J.}} \AND
  \bauthor{\bsnm{Lv},~\bfnm{Jinchi}\binits{J.}}
(\byear{2010}).
\btitle{A selective overview of variable selection in high dimensional feature
  space}.
\bjournal{Statistica Sinica}
\bvolume{20}
\bpages{101-148}.
\end{barticle}
\endbibitem

\bibitem[\protect\citeauthoryear{Fan, Samworth and
  Wu}{2009}]{FanSamworthWu:2009}
\begin{barticle}[author]
\bauthor{\bsnm{Fan},~\bfnm{J.}\binits{J.}},
  \bauthor{\bsnm{Samworth},~\bfnm{R.}\binits{R.}} \AND
  \bauthor{\bsnm{Wu},~\bfnm{Y.}\binits{Y.}}
(\byear{2009}).
\btitle{Ultrahigh dimensional variable selection: beyond the linear model.}
\bjournal{Journal of Machine Learning Research}
\bvolume{10}
\bpages{1829-1853}.
\end{barticle}
\endbibitem

\bibitem[\protect\citeauthoryear{Fan and Song}{2010}]{MMLE2010}
\begin{barticle}[author]
\bauthor{\bsnm{Fan},~\bfnm{J.}\binits{J.}} \AND
  \bauthor{\bsnm{Song},~\bfnm{R.}\binits{R.}}
(\byear{2010}).
\btitle{Sure Independence Screening in Generalized Linear Models With
  NP-Dimensionality}.
\bjournal{Annals of Statistics}
\bvolume{38}
\bpages{3567-3604}.
\end{barticle}
\endbibitem

\bibitem[\protect\citeauthoryear{Friedman, Hastie and
  Tibshirani}{2010}]{glmnetCite}
\begin{barticle}[author]
\bauthor{\bsnm{Friedman},~\bfnm{Jerome}\binits{J.}},
  \bauthor{\bsnm{Hastie},~\bfnm{Trevor}\binits{T.}} \AND
  \bauthor{\bsnm{Tibshirani},~\bfnm{Robert}\binits{R.}}
(\byear{2010}).
\btitle{Regularization Paths for Generalized Linear Models via Coordinate
  Descent.}
\bjournal{Journal of Statistical Software}
\bvolume{33}
\bpages{1-22}.
\end{barticle}
\endbibitem

\bibitem[\protect\citeauthoryear{Horn and Johnson}{2012}]{Horn2012matrix}
\begin{bbook}[author]
\bauthor{\bsnm{Horn},~\bfnm{R.~A.}\binits{R.~A.}} \AND
  \bauthor{\bsnm{Johnson},~\bfnm{C.~R.}\binits{C.~R.}}
(\byear{2012}).
\btitle{Matrix Analysis},
\bedition{2nd} ed.
\bpublisher{Cambridge University Press}.
\end{bbook}
\endbibitem

\bibitem[\protect\citeauthoryear{Hu}{1991}]{SLJHuCum1991}
\begin{barticle}[author]
\bauthor{\bsnm{Hu},~\bfnm{Sau-Lon~James}\binits{S.-L.~J.}}
(\byear{1991}).
\btitle{Probabilistic Independence and Joint Cumulants.}
\bjournal{Journal of Engineering Mechanics}
\bvolume{117}
\bpages{640--652}.
\end{barticle}
\endbibitem

\bibitem[\protect\citeauthoryear{Huang, Li and Wang}{2014}]{Huang2014}
\begin{barticle}[author]
\bauthor{\bsnm{Huang},~\bfnm{Danyang}\binits{D.}},
  \bauthor{\bsnm{Li},~\bfnm{Runze}\binits{R.}} \AND
  \bauthor{\bsnm{Wang},~\bfnm{Hansheng}\binits{H.}}
(\byear{2014}).
\btitle{Feature Screening for Ultrahigh Dimensional Categorical Data with
  Applications}.
\bjournal{Journal of Business and Economic Statistics}
\bvolume{32}
\bpages{237-244}.
\end{barticle}
\endbibitem

\bibitem[\protect\citeauthoryear{Kong, Wang and
  Wahba}{2015}]{KongWangWahba2015}
\begin{barticle}[author]
\bauthor{\bsnm{Kong},~\bfnm{Jing}\binits{J.}},
  \bauthor{\bsnm{Wang},~\bfnm{Sijian}\binits{S.}} \AND
  \bauthor{\bsnm{Wahba},~\bfnm{Grace}\binits{G.}}
(\byear{2015}).
\btitle{Using distance covariance for improved variable selection with
  application to learning genetic risk models}.
\bjournal{Statistics in Medicine}
\bvolume{34}
\bpages{1708--1720}.
\end{barticle}
\endbibitem

\bibitem[\protect\citeauthoryear{Kutner, Nachtsheim and
  Neter}{2004}]{KutnerLinReg2004}
\begin{bbook}[author]
\bauthor{\bsnm{Kutner},~\bfnm{Michael~H}\binits{M.~H.}},
  \bauthor{\bsnm{Nachtsheim},~\bfnm{Chris}\binits{C.}} \AND
  \bauthor{\bsnm{Neter},~\bfnm{John}\binits{J.}}
(\byear{2004}).
\btitle{Applied linear regression models},
\bedition{4th} ed.
\bpublisher{McGraw-Hill/Irwin}.
\end{bbook}
\endbibitem

\bibitem[\protect\citeauthoryear{Li, Zhong and Zhu}{2012}]{DCSIScitation}
\begin{barticle}[author]
\bauthor{\bsnm{Li},~\bfnm{R}\binits{R.}},
  \bauthor{\bsnm{Zhong},~\bfnm{W}\binits{W.}} \AND
  \bauthor{\bsnm{Zhu},~\bfnm{L}\binits{L.}}
(\byear{2012}).
\btitle{Feature Screening via Distance Correlation Learning}.
\bjournal{Journal of the American Statistical Association}
\bvolume{107}
\bpages{1129-1139}.
\end{barticle}
\endbibitem

\bibitem[\protect\citeauthoryear{Liu, Zhong and Li}{2015}]{Liu2015}
\begin{barticle}[author]
\bauthor{\bsnm{Liu},~\bfnm{JingYuan}\binits{J.}},
  \bauthor{\bsnm{Zhong},~\bfnm{Wei}\binits{W.}} \AND
  \bauthor{\bsnm{Li},~\bfnm{RunZe}\binits{R.}}
(\byear{2015}).
\btitle{A selective overview of feature screening for ultrahigh-dimensional
  data}.
\bjournal{Science China Mathematics}
\bvolume{58}
\bpages{1--22}.
\end{barticle}
\endbibitem

\bibitem[\protect\citeauthoryear{Ma, Lan and Wang}{2015}]{Ma2015testing}
\begin{barticle}[author]
\bauthor{\bsnm{Ma},~\bfnm{Yingying}\binits{Y.}},
  \bauthor{\bsnm{Lan},~\bfnm{Wei}\binits{W.}} \AND
  \bauthor{\bsnm{Wang},~\bfnm{Hansheng}\binits{H.}}
(\byear{2015}).
\btitle{Testing predictor significance with ultra high dimensional multivariate
  responses}.
\bjournal{Computational Statistics \& Data Analysis}
\bvolume{83}
\bpages{275--286}.
\end{barticle}
\endbibitem

\bibitem[\protect\citeauthoryear{Nica and Speicher}{2006}]{NicaCum2006}
\begin{bincollection}[author]
\bauthor{\bsnm{Nica},~\bfnm{Alexandru}\binits{A.}} \AND
  \bauthor{\bsnm{Speicher},~\bfnm{Roland}\binits{R.}}
(\byear{2006}).
\btitle{Lectures on the Combinatorics of Free Probability}.
In \bbooktitle{London Mathematical Society Lecture Note Series},
\bvolume{335}
\bpublisher{Cambridge University Press}, \baddress{Oxford, UK}.
\end{bincollection}
\endbibitem

\bibitem[\protect\citeauthoryear{Reese, Dai and Fu}{2018}]{ReeseInteract}
\begin{barticle}[author]
\bauthor{\bsnm{Reese},~\bfnm{Randall}\binits{R.}},
  \bauthor{\bsnm{Dai},~\bfnm{Xiaotian}\binits{X.}} \AND
  \bauthor{\bsnm{Fu},~\bfnm{Guifang}\binits{G.}}
(\byear{2018}).
\btitle{Strong Sure Screening of Ultrahigh Dimensional Feature Spaces with
  Interaction Effects}.
\bjournal{To Appear}.
\end{barticle}
\endbibitem

\bibitem[\protect\citeauthoryear{Seghouane}{2011}]{Seghouane2011new}
\begin{barticle}[author]
\bauthor{\bsnm{Seghouane},~\bfnm{Abd-Krim}\binits{A.-K.}}
(\byear{2011}).
\btitle{New AIC corrected variants for multivariate linear regression model
  selection}.
\bjournal{IEEE Transactions on Aerospace and Electronic Systems}
\bvolume{47}
\bpages{1154--1165}.
\end{barticle}
\endbibitem

\bibitem[\protect\citeauthoryear{Serfling}{1980}]{Serfling1980}
\begin{bbook}[author]
\bauthor{\bsnm{Serfling},~\bfnm{R.~J.}\binits{R.~J.}}
(\byear{1980}).
\btitle{Approximation Theorems of Mathematical Statistics}.
\bpublisher{Wiley}, \baddress{New York}.
\end{bbook}
\endbibitem

\bibitem[\protect\citeauthoryear{Sun, Peng and Shakoor}{2014}]{Sun2014kernel}
\begin{barticle}[author]
\bauthor{\bsnm{Sun},~\bfnm{Shiquan}\binits{S.}},
  \bauthor{\bsnm{Peng},~\bfnm{Qinke}\binits{Q.}} \AND
  \bauthor{\bsnm{Shakoor},~\bfnm{Adnan}\binits{A.}}
(\byear{2014}).
\btitle{A kernel-based multivariate feature selection method for microarray
  data classification}.
\bjournal{PloS one}
\bvolume{9}
\bpages{e102541}.
\end{barticle}
\endbibitem

\bibitem[\protect\citeauthoryear{Tibshirani}{1996}]{Tibshirani}
\begin{barticle}[author]
\bauthor{\bsnm{Tibshirani},~\bfnm{R.}\binits{R.}}
(\byear{1996}).
\btitle{Regression shrinkage and selection via lasso}.
\bjournal{Journal of the Royal Statistical Society: Series B}
\bvolume{58}
\bpages{267-288}.
\end{barticle}
\endbibitem

\bibitem[\protect\citeauthoryear{van Buuren and
  Groothuis-Oudshoorn}{2011}]{JSSv045i03}
\begin{barticle}[author]
\bauthor{\bparticle{van} \bsnm{Buuren},~\bfnm{Stef}\binits{S.}} \AND
  \bauthor{\bsnm{Groothuis-Oudshoorn},~\bfnm{Karin}\binits{K.}}
(\byear{2011}).
\btitle{mice: Multivariate Imputation by Chained Equations in R}.
\bjournal{Journal of Statistical Software, Articles}
\bvolume{45}
\bpages{1--67}.
\bdoi{10.18637/jss.v045.i03}
\end{barticle}
\endbibitem

\bibitem[\protect\citeauthoryear{Zhang et~al.}{2012}]{Zhang2012Mice}
\begin{barticle}[author]
\bauthor{\bsnm{Zhang},~\bfnm{Weidong}\binits{W.}},
  \bauthor{\bsnm{Korstanje},~\bfnm{Ron}\binits{R.}},
  \bauthor{\bsnm{Thaisz},~\bfnm{Jill}\binits{J.}},
  \bauthor{\bsnm{Staedtler},~\bfnm{Frank}\binits{F.}},
  \bauthor{\bsnm{Harttman},~\bfnm{Nicole}\binits{N.}},
  \bauthor{\bsnm{Xu},~\bfnm{Lingfei}\binits{L.}},
  \bauthor{\bsnm{Feng},~\bfnm{Minjie}\binits{M.}},
  \bauthor{\bsnm{Yanas},~\bfnm{Liane}\binits{L.}},
  \bauthor{\bsnm{Yang},~\bfnm{Hyuna}\binits{H.}},
  \bauthor{\bsnm{Valdar},~\bfnm{William}\binits{W.}},
  \bauthor{\bsnm{Churchill},~\bfnm{Gary~A.}\binits{G.~A.}} \AND
  \bauthor{\bsnm{DiPetrillo},~\bfnm{Keith}\binits{K.}}
(\byear{2012}).
\btitle{Genome-Wide Association Mapping of Quantitative Traits in Outbred
  Mice}.
\bjournal{G3: Genes, Genomes, Genetics}
\bvolume{2}
\bpages{167--174}.
\bdoi{10.1534/g3.111.001792}
\end{barticle}
\endbibitem

\bibitem[\protect\citeauthoryear{Zhao and Li}{2012}]{ZHAO2012397}
\begin{barticle}[author]
\bauthor{\bsnm{Zhao},~\bfnm{Sihai~Dave}\binits{S.~D.}} \AND
  \bauthor{\bsnm{Li},~\bfnm{Yi}\binits{Y.}}
(\byear{2012}).
\btitle{Principled sure independence screening for Cox models with
  ultra-high-dimensional covariates}.
\bjournal{Journal of Multivariate Analysis}
\bvolume{105}
\bpages{397 - 411}.
\bdoi{https://doi.org/10.1016/j.jmva.2011.08.002}
\end{barticle}
\endbibitem

\bibitem[\protect\citeauthoryear{Zhong and Zhu}{2014}]{ZhongZhu2014}
\begin{barticle}[author]
\bauthor{\bsnm{Zhong},~\bfnm{W}\binits{W.}} \AND
  \bauthor{\bsnm{Zhu},~\bfnm{L}\binits{L.}}
(\byear{2014}).
\btitle{An Iterative Approach to Distance Correlation-Based Sure Independent
  Screening}.
\bjournal{Journal of Statistical Computation and Simulation}
\bpages{1-15}.
\end{barticle}
\endbibitem

\bibitem[\protect\citeauthoryear{Zhu et~al.}{2011}]{ZhuLiLiZhu2011}
\begin{barticle}[author]
\bauthor{\bsnm{Zhu},~\bfnm{Li-Ping}\binits{L.-P.}},
  \bauthor{\bsnm{Li},~\bfnm{Lexin}\binits{L.}},
  \bauthor{\bsnm{Li},~\bfnm{Runze}\binits{R.}} \AND
  \bauthor{\bsnm{Zhu},~\bfnm{Li-Xing}\binits{L.-X.}}
(\byear{2011}).
\btitle{Model-Free Feature Screening for Ultrahigh Dimensional Data}.
\bjournal{Journal of the American Statistical Association}
\bvolume{106}
\bpages{1464-1475}.
\end{barticle}
\endbibitem

\bibitem[\protect\citeauthoryear{Zou}{2006}]{ZouAdaptLasso2006}
\begin{barticle}[author]
\bauthor{\bsnm{Zou},~\bfnm{Hui}\binits{H.}}
(\byear{2006}).
\btitle{The adaptive lasso and its oracle properties}.
\bjournal{Journal of the American Statistical Association}
\bvolume{101}
\bpages{1418-1429}.
\end{barticle}
\endbibitem

\bibitem[\protect\citeauthoryear{Zou and Hastie}{2005}]{elasticNet}
\begin{barticle}[author]
\bauthor{\bsnm{Zou},~\bfnm{Hui}\binits{H.}} \AND
  \bauthor{\bsnm{Hastie},~\bfnm{Trevor}\binits{T.}}
(\byear{2005}).
\btitle{Regularization and variable selection via the elastic net}.
\bjournal{Journal of the Royal Statistical Society, Series B}
\bvolume{67}
\bpages{301-320}.
\end{barticle}
\endbibitem

\end{thebibliography}

\end{document}